\UseRawInputEncoding

\documentclass[
preprint, 
preprintnumbers,superscriptaddress,showpacs,supberscriptaddress,
nofootinbib,longbibliography]{revtex4-1}

\usepackage[
bookmarks=true,colorlinks,linkcolor=blue,urlcolor=cyan,citecolor=red]{hyperref}

\usepackage{graphicx}
\usepackage{latexsym}
\usepackage{amsfonts}
\usepackage{amssymb}
\usepackage{amsmath}
\usepackage{bm}
\usepackage{mathrsfs}
\usepackage{slashed}
\usepackage{ascmac}
\usepackage{comment}

\usepackage{dcolumn}

\allowdisplaybreaks[3]

\usepackage[
bookmarks=true,colorlinks,linkcolor=blue,urlcolor=cyan,citecolor=red]{hyperref}

\newcommand{\cout}[1]{ \if 0 {#1} \fi }

\newcommand{\beq}{\begin{eqnarray}}
\newcommand{\eeq}{\end{eqnarray}}
\newcommand{\bseq}{\begin{subequations}}
\newcommand{\eseq}{\end{subequations}}
\newcommand{\nn}{\nonumber}
\renewcommand{\=}{&=&}
\newcommand{\nnb}{\nonumber \\}
\newcommand{\pd}{\partial}

\newcommand{\sla}{ \slashed }

\newcommand{\ep}{ \epsilon }

\newcommand{\gam}{ \gamma }

\newcommand{\bx}{{\bm{x}}}

\newcommand{\bp}{{\bm{p}}}
\newcommand{\bq}{{\bm{q}}}

\newcommand{\bB}{{\bm B}}
\newcommand{\bE}{{\bm E}}

\newcommand{\order}{{\mathcal O}}
\newcommand{\prj}{ {\mathcal P} }

\newcommand{\para}{ \parallel}

\newcommand{\tr}{ {\rm tr} }
\newcommand{\diag}{ {\rm diag} }
\newcommand{\sgn}{ {\rm sgn} }

\newcommand{\vac}{ {\rm vac} }

\newcommand{\LLL}{ {\rm LLL} }

\usepackage{color} 
\usepackage[normalem]{ulem}  

\newcommand{\temp}{{\rm med}}

\begin{document}

\title{In-medium polarization tensor in strong magnetic fields (II): 
\\
Axial Ward identity at finite temperature and density}

\author{Koichi Hattori}
\affiliation{Zhejiang Institute of Modern Physics, Department of Physics, 
Zhejiang University, Hangzhou, 310027, China} 
\affiliation{Research Center for Nuclear Physics, Osaka University, Osaka 567-0047 Japan.}

\author{Kazunori Itakura}
\affiliation{Nagasaki Institute of Applied Science (NiAS), 536 Aba-machi, 
Nagasaki-shi, Nagasaki, 851-0193, JAPAN}
\affiliation{High-Energy Accelerator Research Organization (KEK), 1-1 Oho, 
Tsukuba, Ibaraki, 305-0801, JAPAN}


\begin{abstract} 
We investigate the axial Ward identity (AWI) for massive fermions in strong magnetic fields. 
The divergence of the axial-vector current is computed at finite temperature and/or density 
with the help of a relation between the polarization and anomaly diagrams 
in the effective (1+1) dimensions realized in the lowest Landau level (LLL). 
We discuss delicate interplay between the vacuum and medium contributions 
that determines patterns of the spectral flow in the adiabatic limit 
and, more generally, the diabatic chirality production rate. 
We also establish an explicit relation between the AWIs from the LLL approximation 
and from the familiar triangle diagrams in the naive perturbative series 
with respect to the coupling constant.


\end{abstract}

\maketitle
\tableofcontents

\section{Introduction}

In this series of papers, we investigate the polarization effects of the medium particles 
at finite temperature and/or density as well as of the vacuum fluctuations in strong magnetic fields. 
In the first paper \cite{Hattori:2022uzp}, 
which is referred to as Paper I, we have provided the explicit forms of 
the vacuum and medium contributions to the one-loop polarization tensor 
in the lowest Landau level (LLL) approximation and discussed magneto-birefringence 
as a physical application, that is, the polarization-dependent photon dispersion relation 
induced by the strong magnetic fields. 
We have inspected delicate interplay between the vacuum and medium contributions.

In the second paper of the series, we discuss another application 
of the polarization tensor to the axial-charge generation. 
Although the polarization tensor is a current correlator of the vector-vector type, 
one can apply it to the induction of the axial-vector current because there is an identity between 
the vector and axial-vector currents expanded with the eigenstates of the Landau levels. 
This identity is an analogue of that in the (1+1)-dimensional QED 
and the physical reason behind this identity is the chirality-momentum locking 
in the spin eigenstates (see Sec.~\ref{sec:VA} for more details).

Mechanisms of the axial-charge generation have attracted much attention over the last decade 
since a nonzero axial charge implies a parity-odd environment 
and induces transport phenomena that do not occur in parity-even environments. 
One of the prominent examples is induction of an electric current in magnetic fields, 
which is known as the chiral magnetic effect (CME) \cite{Kharzeev:2007jp, Fukushima:2008xe}. 
The anomalous transport phenomena have been intensively investigated 
in the contexts of quark-gluon plasma 
created by relativistic heavy-ion collisions \cite{Kharzeev:2015znc, Skokov:2016yrj, Hattori:2016emy} 
and of the Weyl and Dirac semimetals in condensed matter physics 
\cite{Burkov:2015hba, Gorbar:2017lnp, 2018RvMP...90a5001A}. 
The chiral magnetic effect, together with Amp\`ere's law, 
further implies an exponential growth of the magnetic-field strength 
driven by a positive feedback cycle when the magnetic field is lifted to a dynamical field. 
This is called the chiral plasma instability \cite{Akamatsu:2013pjd} 
and is expected to be possible mechanisms for generation of 
the strong magnetic fields during supernova explosions \cite{Yamamoto:2015gzz, Yamamoto:2020zrs} 
and of the primordial magnetic field in the early universe \cite{Turner:1987bw, 
Carroll:1989vb, Garretson:1992vt, Joyce:1997uy, Giovannini:2003yn, Laine:2005bt}. 
Those hot situations, including the very recent CME search by isobaric collisions \cite{STAR:2021mii}, 
motivate us to investigate the axial-charge generation. 
Although the CME conductivity is given by a universal form stemming from chiral anomaly, 
it is also proportional to the magnitude of the axial chemical potential 
that is a thermodynamic conjugate quantity to the axial-charge density 
and depends on dynamics in individual systems. 
To name a few, the axial-charge dynamics was investigated in 
Refs.~\cite{Tanji:2016dka, Mueller:2016ven,Hou:2017szz, Tanji:2018qws, 
Copinger:2018ftr, Taya:2020bcd, Muller:2021hpe, Yang:2021fea} for relativistic heavy-ion collisions, 
Refs.~\cite{Grabowska:2014efa, Gorbar:2021tnw} for neutron-star physics,  
and Refs.~\cite{Domcke:2019qmm, Boyarsky:2020cyk} for cosmology.

We investigate the axial-charge generation from a fundamental point of view 
in terms of the divergence of the axial-vector current.  
QED contains chiral anomaly \cite{Adler:1969gk, Bell:1969ts}, 
giving rise to the anomalous nonvanishing term in the divergence of the axial-vector current.  
A finite fermion mass explicitly breaks the chiral symmetry, and 
gives rise to another nonvanishing term in the divergence of the current. 
In case of (3+1)-dimensional QED, the well-known triangle diagrams 
yield both of those two nonvanishing contributions 
when the loop diagrams are composed of massive fermions.

There is, however, a significant difference between those two pieces of the same diagrams. 
The anomalous term does not receive radiative corrections \cite{Adler:1969er}; 
The anomalous term stems from the (superficial) ultraviolet divergence of the triangle diagrams, 
so that, without modifications of the divergent pole, there is no correction to the anomalous term. 
This means that there is no correction from 
the fermion mass or other external parameters such as temperature or density 
that do not modify the divergent pole \cite{Itoyama:1982up}. 
In contrast, the mass-dependent part does not involve an ultraviolet divergence 
and can come with temperature and/or density corrections. 
The whole non-conservation equation of the axial-vector current, 
which we call the axial Ward identity (AWI), is controlled by 
the balance between the anomalous and mass-dependent terms 
that arise from the divergent and finite pieces of the loop integral.\footnote{
The mass-dependent part may not be saturated with the finite piece of the anomaly diagrams, 
and could be subject to higher-order or any kind of corrections due to the non-divergent nature.}

Since the mass-dependent part in the AWI is not anomalous in nature, 
it does not exhibit simple behaviors in general when one varies magnitudes of 
relevant parameters such as the external photon momenta, fermion mass, temperature and/or density. 
Therefore, the AWI as a whole is not simply governed by the anomalous term 
and, moreover, the two terms can be comparable in magnitude, 
implying that the axial-vector current could be effectively conserved in some parameter regimes. 
Thus, we investigate the whole AWI with explicit computations.

Relevant processes for the axial-charge generation can be classified 
into {\it adiabatic} and {\it diabatic} processes according to 
the magnitude of energy transfer from the external photon fields 
(see Sec.~\ref{sec:brief-summary} for a more precise definition). 
The adiabatic processes can be understood as the spectral flow that is  
a collective flow of the occupied states along the fermion dispersion relation 
\cite{Nielsen:1983rb, Ambjorn:1983hp} (see also Ref.~\cite{Creutz:2000bs} for a review). 
The adiabatic flows occur when there is no energy transfer from the photon fields to the fermions 
and are only the processes for which one can get simple results. 
One should notice that the spectral flow occurs on the thermally occupied positive-energy states 
as well as on the negative-energy states beneath the Dirac sea, 
and induces the helicity flip at the bottom of the parabolic dispersion relation for massive fermions. 
In the other cases, i.e., the diabatic processes, 
nonzero energy transfer can induce intricate excitations that ruin the ordered collective flows. 
Such information is all encoded in the mass-dependent part. 
We provide a more detailed summary of the results in Sec.~\ref{sec:brief-summary}.

Incidentally, we also explicitly show a connection between the AWIs 
from the triangle diagrams and from the polarization tensor in the LLL approximation. 
Chiral anomaly is often explained as a product of the (1+1)-dimensional chiral anomaly 
and the Landau degeneracy factor on the basis of the effective dimensional reduction 
in the LLL \cite{Nielsen:1983rb}. 
The explicit computation of the LLL contribution, however, does not exactly reproduce 
the result from the triangle diagrams due to an extra Gaussian factor 
that stems from the fermion wave function in the LLL. 
This is simply because the wave function of the LLL fermion is not exactly 
(1+1) dimensional one, but has an extension in perpendicular to the magnetic field. 
The Gaussian factor reduces to unity in the homogeneous electric-field limit 
and/or the strong magnetic field limit such that 
the electric field does not resolve the transverse extension of the cyclotron orbits. 
As for the mass-dependent part, the LLL approximation reproduces that from the triangle diagrams 
when the {\it constant} magnetic field limit is taken for the triangle diagrams, 
again up to the Gaussian factor.

This paper is organized as follows. 
In Sec.~\ref{sec:VA}, we provide basic formulas for the induction 
of the vector and axial-vector currents by a single two-point correlator in the LLL approximation. 
In Sec.~\ref{sec:polarization-tensor}, we summarize the explicit expressions 
of the polarization tensor from Paper I for readers' convenience. 
Then, we discuss the AWI in detail in Sec.~\ref{sec:AWI}. 
In appendices, we clarify the quantum numbers characterizing the fermions in the LLL 
and perform an explicit computation of the divergence of the axial current 
composed of a massive Dirac fermion in the LLL, reproducing the expression from the polarization tensor. 
Finally, we discuss a connection between the AWIs computed from the LLL approximation 
and the triangle diagrams in Appendix~\ref{sec:triangle-LLL}.

As in Paper I, we focus on a single-flavor Dirac fermion with an electric charge $ q_f $ for simplicity, 
and introduce a single chemical potential conjugate to its number density. 
We also assume that a constant magnetic field is applied along the third spatial direction 
without loosing generality, i.e., $ \bB = (0,0,B) $. 
Accordingly, we use metric conventions $ g^{\mu\nu} = \diag ( 1, -1, -1, -1) $, 
$ g_\para^{\mu\nu} = \diag(1, 0,0,-1) $, and $ g_\perp^{\mu\nu} = (0,-1,-1,0) $ 
and associated notations $ q_\para^2 := q_\mu q_\nu g_\para^{\mu\nu} $ 
and $ q_\perp^2 := q_\mu q_\nu g_\perp^{\mu\nu} = - |\bq_\perp|^2$ for four vectors $ q^\mu $.

\section{Vector and axial Ward identities}

\label{sec:VA}

The retarded polarization tensor serves as a linear response function 
with respect to electromagnetic perturbations. 
In this paper, we discuss induction of the currents transported by the LLL fermions 
\begin{subequations}
\begin{eqnarray}
\label{eq:jV-def}
j_V^\mu \= q_f \bar \psi_\LLL \gam^\mu \psi_\LLL
\, ,
\\
\label{eq:jA-def}
j_A^\mu \= \bar \psi_\LLL  \gam^\mu \gam^5 \psi_\LLL
\, .
\end{eqnarray}
\end{subequations}
The fermion spinor $ \psi_\LLL $ in the LLL is an eigenstate of the spin projection operator 
$  \prj_+ = (1+ is_f \gam^1 \gam^2)/2$ with the sign function $ s_f  =\sgn(q_f B) $. 
By the use of an identity among the gamma matrices $ 
\gam_\para^\mu \gam^5 \prj_\pm = \mp s_f \epsilon _\para^{\mu \nu} \gam_{\para \nu} \prj_\pm$, 
we find a relation between the LLL contributions to 
the vector and axial-vector currents along the magnetic field\footnote{
Here, an electric charge is not included in the definition of 
the axial-vector current (\ref{eq:jA-def}), while it is included in the vector current (\ref{eq:jV-def}) 
and the electromagnetic coupling $j_{V}^\mu A_\mu  $ in Eq.~(\ref{eq:anomaly-LLL-1}). 
} 
\begin{eqnarray}
q_f j_A^\mu = - s_f  \epsilon _\para^{\mu \nu} j_{V \nu}
\label{eq:V-A}
\, ,
\end{eqnarray}
where we introduced an antisymmetric tensor $  \epsilon _\para^{\mu \nu}  $ 
which only has two non-vanishing components, $ \epsilon_\para^{03} = -\epsilon_\para^{30} =1 $. 
This relation is an analogue of the well-known relation 
in the (1+1) dimensional QED (see, e.g., a standard textbook \cite{Peskin:1995ev})
and is understood as a consequence of the effective dimensional reduction.  
Equation (\ref{eq:V-A}) relates the vector--vector (VV) correlator $ \Pi^{\mu\nu}_R $, 
that we inspected in Paper I, to the vector--axial-vector (VA) correlator 
\begin{eqnarray}
\langle j_A^\mu j_V^\nu \rangle_R = - s_f  \epsilon _{\para \, \rho}^{\mu} q_f^{-1} \Pi^{\rho\nu}_R  
\, .
\end{eqnarray}

The above vector and axial-vector currents are induced in response 
to an external $U(1)$ gauge field $ A^\mu $. 
This is an external gauge field with an arbitrary frequency and wavelength, 
superimposed on top of the constant magnetic field. 
When the polarization tensor has a gauge-invariant tensor structure 
[see Eq.~(\ref{eq:transverse-op})], 
the vector and axial Ward identities read 
\begin{subequations}
\label{eq:Ward-ids}
\begin{eqnarray}
&&
{\rm F.T.} \ \pd_\mu j_V^\mu =  -i q_\mu  \Pi^{\mu\nu}_R   A_\nu =0
\label{eq:Ward-id1}
\, ,
\\
&&
{\rm F.T.} \ \pd_\mu j_A^\mu = -i q_\mu \langle j_A^\mu j_V^\nu \rangle_R A_\nu 
=  s_f  q_f^{-1} \Pi_\para \tilde E  _\para (q)
\label{eq:anomaly-LLL-1}
\, ,
\end{eqnarray}
\end{subequations}
where F.T. stands for the Fourier transform 
and $ \tilde E  _\para (q) = - i \epsilon _\para^{\mu \nu} q_\mu A_\nu $ is the parallel (or antiparallel) 
component of an electric field along the magnetic field. 
We put tilde on the electric field to emphasize that it is the Fourier spectrum, 
while the magnetic field $ B $ (and $ \bB $ appearing below) is 
the constant magnitude throughout this paper. 
The vector Ward identity (\ref{eq:Ward-id1}) is a consequence of the gauge invariance in QED. 
However, the axial current is not conserved if the right-hand side 
in Eq.~(\ref{eq:anomaly-LLL-1}) is nonzero. 
The magnitude of the nonvanishing divergence is controlled by 
the scalar function $  \Pi_\para $ from the polarization tensor. 
In the next section, we provide the explicit form of $  \Pi_\para $ 
from the vacuum and medium contributions.

\section{Polarization tensor}

\label{sec:polarization-tensor}

\begin{figure}
     \begin{center}
              \includegraphics[width=0.8\hsize]{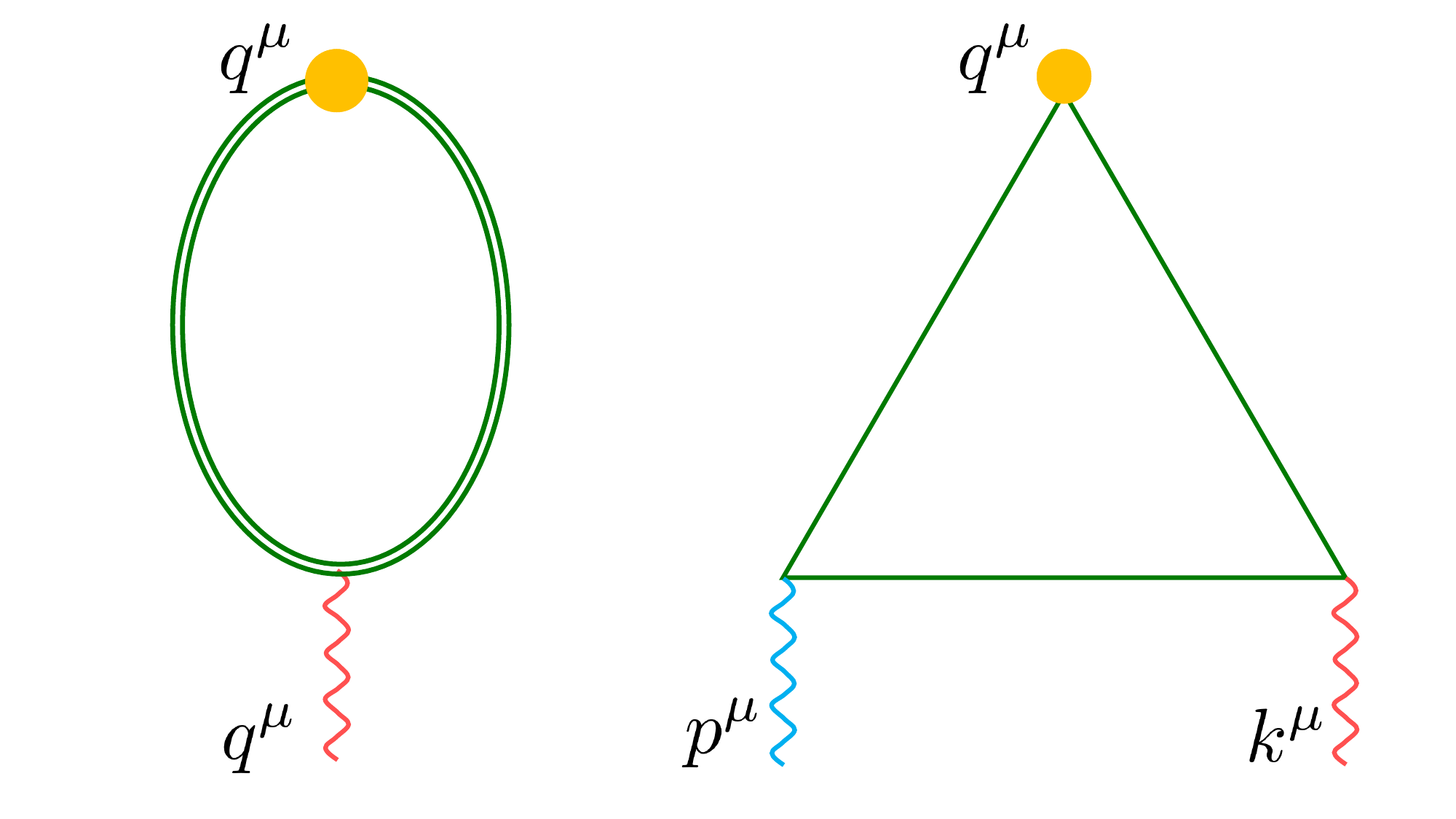}
     \end{center}
\caption{One-loop polarization tensor in strong magnetic fields [left] 
and the triangle diagram (with an exchange diagram not shown) [right]. 
The double lines show the fermion propagators in the LLL approximation 
that have insertions of the magnetic-field lines (not shown).}
\label{fig:diagrams}
\end{figure}

In Paper I, we provided detailed accounts on the computation of 
the polarization tensor in the LLL approximation (cf. the left panel in Fig.~\ref{fig:diagrams}). 
Here, we briefly summarize its explicit form for readers' convenience.

In the LLL approximation, one obtains the polarization tensor 
\begin{eqnarray}
\Pi^{\mu\nu} = m_B^2 e^{-\frac{\vert \bq_\perp \vert^2}{2 \vert q_fB\vert}} \tilde  \Pi_\para \prj_\para^{\mu\nu} 
\label{eq:transverse-op}
\, ,
\end{eqnarray}
where $m_B^2 = |q_fB|/(2\pi) \cdot q_f^2/\pi  $. 
The gauge-invariant tensor structure is given as 
\begin{eqnarray}
\label{eq:P-para}
 \prj_\para^{\mu\nu}  =  g^{\mu\nu}_\para - \frac{ q_\para^\mu q_\para^\nu}{q_\para^2} 
 \, ,
\end{eqnarray}
with the photon momentum $ q^\mu $. 
Note that this tensor structure cannot be split into the longitudinal and transverse 
components with respect to the spatial photon momentum, 
because of its dimensionally reduced form. 
Therefore, this is the unique gauge-invariant tensor structure within the LLL approximation. 
There are some other tensor structures when the higher Landau levels contribute 
(see, e.g., Refs.~\cite{Hattori:2012je, Hattori:2017xoo}). 
The polarization function $  \tilde  \Pi_\para $ contains both the vacuum and medium contributions 
\begin{eqnarray}
\tilde \Pi_\para  = \tilde \Pi_\para ^\vac (q_\para^2) + \tilde \Pi_\para ^\temp(\omega, q_z)
\, .
\end{eqnarray}
The medium contribution $\tilde \Pi_\para ^\temp(\omega, q_z) $ depends on 
the photon energy $ \omega $ and momentum $ q_z $ separately, 
because the presence of medium breaks the Lorentz-boost invariance. 
Note also that we apply a constant magnetic field $ \bB $ in the medium rest frame. 
Configurations composed of constant magnetic fields in other Lorentz frames 
are not equivalent to the present configuration.

The explicit diagram computation provides the vacuum contribution (see the appendix in Paper I) 
\begin{eqnarray}
\tilde  \Pi_\para^\vac  (q_\para^2) =   1- I \big( \frac{q_\parallel^2}{4m^2} \big) 
\label{eq:vac_massive}
\, .
\end{eqnarray}
Since the overall factor of $ m_B^2  $ is a dimension-two quantity in Eq.~(\ref{eq:transverse-op}), 
the $ I $ function can only be a function of a dimensionless variable: 
\begin{eqnarray}
I( x + i \omega \epsilon ) = 
\left\{
\begin{array}{ll}
\frac{1}{2} \frac{1}{ \sqrt{x(x-1) } }
\ln  \frac{ \sqrt{x(x-1) } - x }{  \sqrt{x(x-1) } + x } 
& x < 0 
\\
\frac{1}{ \sqrt{x(1 - x) } }\arctan \frac{x } { \sqrt{x(1-x) } } 
& 0 \leq x < 1
\\
\frac{1}{2} \frac{1}{ \sqrt{x(x-1) } }
\left[ \,\ln  \frac{x -  \sqrt{x(x-1) } }{x +  \sqrt{x(x-1) } } 
+ i \,  \sgn(\omega)  \pi \, \right]
& 1 \leq x
\end{array}
\right.
\label{eq:I}
\, .
\end{eqnarray}  
We took the retarded prescription $ \omega \to \omega + i\epsilon $ that 
yields an infinitesimal displacement $ q_\para^2 \to q_\para^2 + i \omega \epsilon $. 
The advanced and causal correlators can be computed in the same manner.  
One can find two limiting behaviors 
\begin{eqnarray}
\label{eq:I-limits}
I(0) = 1 \, , \quad I(\infty) = 0
\, .
\end{eqnarray}
Therefore, in the infinite-mass limit $  q_\para^2/m^2 \to 0$, 
we simply find that $\tilde  \Pi_\para^\vac  \to 0 $ 
because the fermion excitations are suppressed. 
In the massless limit $  q_\para^2/m^2 \to \infty$, we have
\begin{eqnarray}
\tilde  \Pi_\para^\vac  (q_\para^2) \=   1  
\label{eq:Pi-massless}
\, .
\end{eqnarray} 
The factor of $ |q_fB|/(2\pi)  $ in $m_B^2  $ is called the Landau degeneracy factor. 
The remaining factor $ q_f^2/\pi  $ is known as the Schwinger mass 
in the (1+1) dimensional massless QED \cite{Schwinger:1962tn, Schwinger:1962tp} 
that provides a photon mass in the gauge-invariant way.


The medium contribution is obtained as (see Ref.~\cite{Fukushima:2015wck} and Paper I) 
\begin{eqnarray}
\tilde \Pi_\para^{\temp }(\omega, q_z) 
=  \pi m^2 \!\! \int_{-\infty}^\infty \frac{d p_z}{2\pi \epsilon_p} 
\frac{ ( q_\parallel^2 + 2 q_z p_z) \,   [n_+(\epsilon_p)  + n_-(\epsilon_p)] } 
{  q_\parallel^2  ( p_z - \frac{1}{2} q_z) ^2 - \frac{\omega^2}{4 }  (q_\parallel^2 -4m^2) }
\label{eq:Pi_medium}
\, ,
\end{eqnarray}
where the fermion energy and distribution function are given as 
$ \epsilon_p =\sqrt{p_z^2 + m^2} $ and 
$ n_\pm(\epsilon_p) = [ e^{ (\epsilon_p\mp\mu)/T} + 1]^{-1} $, respectively. 
Notice that the medium contribution is proportional to the fermion mass 
and does not exist in the strictly massless case. 
The integrand has two poles at 
\begin{eqnarray}
\label{eq:pair-momentum}
 p^{\pm}_z  := \frac{1}{2} \big( \, q_z \pm \omega \sqrt{1- 4m^2/q_\para^2 } \ \big)
 \, .
\end{eqnarray} 
Thus, the integrand in Eq.~(\ref{eq:Pi_medium}) can be rearranged as 
\begin{eqnarray}
\tilde \Pi_\para^{\temp }(\omega, q_z) 
=   \frac{ 2 \pi  m^2  }{ q_\para^2\sqrt{1- 4m^2/q_\para^2 }  }
 \int_{-\infty}^\infty \frac{d p_z}{2\pi \epsilon_p} 
[ n_+(\epsilon_p)  + n_-(\epsilon_p)]
\Big( \frac{ \epsilon_p^+ s_+ }{ p_z - p_z^+ } - \frac{ \epsilon_p^- s_-}{ p_z - p_z^- } \Big)
\label{eq:Pi_medium-1}
\, ,
\end{eqnarray}
where we defined 
\begin{subequations}
\label{eq1}
\begin{eqnarray}
\label{eq:energy-signs}
&&s_\pm = \sgn \Big(\omega \pm q_z    \sqrt{1- 4m^2/q_\para^2 } \Big)
= \sgn(\omega) \,  \theta(q_\para^2 -4m^2) \pm \theta(-q_\para^2)
\, ,
\\
\label{eq:pair-energy}
&&\ep_p^\pm = \sqrt{ (p^{\pm}_z)^2 + m^2 }
= \frac{1}{2} \Big \vert  \omega \pm  q_z   \sqrt{ 1 -  4m^2/q_\para^2 } \ \Big \vert
\, ,
\end{eqnarray}
\end{subequations}
for $  1 -  4m^2/q_\para^2 \geq 0 $ and 
\begin{subequations}
\label{eq2}
\begin{eqnarray}
\label{eq:energy-signs-2}
&&s_\pm = 1
\, ,
\\
\label{eq:pair-energy-2}
&&\ep_p^\pm = \frac{1}{2} \Big (  \omega \pm  i q_z   \sqrt{ |1 -  4m^2/q_\para^2| } \ \Big )
\, ,
\end{eqnarray}
\end{subequations}
for $ 1 -  4m^2/q_\para^2 <0  $.

Lastly, we give some comments on the imaginary part of the polarization tensor. 
The imaginary part in Eq.~(\ref{eq:I}) indicates occurrence of 
a fermion-antifermion pair creation when the photon momentum satisfies 
the threshold condition $ q_\para^2 \geq (2m)^2 $ 
that is the invariant mass of a pair in the LLL states (cf. Paper I). 
In the massless limit (\ref{eq:Pi-massless}), 
an imaginary part only comes from the $ q_\para^2 $ pole in Eq.~(\ref{eq:P-para}).\footnote{
This pole does not exist in the massive case according to the limit (\ref{eq:I-limits}).
} 
The imaginary part is thus proportional to $ \delta(q_\para^2) $, meaning that 
there is no invariant energy transfer from a photon to a fermion-antifermion pair. 
Such a pair creation does not usually happen in (3+1) dimensions 
since a hole state in the Dirac sea needs to be 
created by a {\it diabatic} transition of a particle from the negative- to positive-energy branches. 
The pair creation in an {\it adiabatic} process can occur only because 
the positive- and negative-energy states are directly connected with each other 
in the (1+1)-dimensional gapless dispersion relation.
Namely, the massless pair creation here can be interpreted as the spectral flow 
along the linear dispersion relation that occurs independently in the right- and left-handed sectors, 
leading to chiral anomaly. 
There is no medium correction in the massless case. 
The medium contribution (\ref{eq:Pi_medium-1}) acquires imaginary parts 
in the massive case when the fermion and antifermion pair takes the energy $ \ep_p^\pm $ 
and momentum $  p^{\pm}_z $ as we inspected in Paper I. 
The created massive pairs posses nonzero axial charges.


\cout{

By the use of the familiar formula 
\begin{eqnarray}
\label{eq:formula1}
1/(x \pm i \epsilon) = P(1/x) \mp i\pi \delta (x) 
\, ,
\end{eqnarray}
one can extract the imaginary part of the polarization tensor as 
\begin{eqnarray}
\Im m \tilde   \Pi_\para^{\temp } 
\=  \frac{ 2 \pi m_B^2 \, m^2 } {   q_\parallel^2    \sqrt{ 1-4m^2/q_\para^2}  } 
\left[   \, s_+  \frac{n_+(\epsilon_p^+)  +  n_- (\epsilon_p^+)}{2}  
+  s_-   \frac{n_+(\epsilon_p^-)  +  n_- (\epsilon_p^-)}{2}   \, \right]
\label{eq:ImPi-med}
\, .
\end{eqnarray}

Lastly, the imaginary part in Eq.~(\ref{eq:I}) indicates occurrence of 
a fermion-antifermion pair creation when the photon momentum satisfies 
the threshold condition $ q_\para^2 \geq (2m)^2 $ that is the invariant mass of a pair in the LLL states. 
In the massless limit (\ref{eq:Pi-massless}), an imaginary part only comes from the $ q_\para^2 $ pole. 
The imaginary part is proportional to $ \delta(q_\para^2) $, meaning that 
there is no invariant momentum transfer from a photon to a fermion-antifermion pair. 
Such a pair creation does not usually happen since a hole state in the Dirac sea needs to be 
created by a {\it diabatic} transition of a particle from the negative- to positive-energy branches. 
The pair creation in an {\it adiabatic} process can occur only because 
the positive- and negative-energy states are directly connected with each other 
in the (1+1)-dimensional gapless dispersion relation.\footnote{
On the other hand, the diabatic pair creation is prohibited in the massless case 
due to the absence of chirality mixing. 
If a fermion and antifermion pair were created, they should have the same helicity 
in the center-of-momentum frame of the pair, meaning that they belong to 
different chirality eigenstates in strict massless theories. 
A similar prohibition mechanism is known as the ``helicity suppression'' 
in the leptonic decay of charged pions \cite{Donoghue:1992dd, Zyla:2020zbs}. 
} 
Namely, the pair creation is interpreted as the spectral flow along the linear dispersion relation 
that occurs independently in the right- and left-handed sectors, leading to the chiral anomaly. 
Finite mass effects, which are captured by the $ I $ function, modify the mechanism of the pair creation. 
We will discuss the AWI that contains not only the nonrenormalizable anomalous term 
but also the pseudoscalar condensate term. 
The latter depends on the fermion mass as well as temperature/density, 
so that the AWI also depend on those parameters as a whole. 

}

\section{Axial Ward identity in strong magnetic fields}

\label{sec:AWI}

\subsection{Summary of the results}

\label{sec:brief-summary}

\cout{

\begin{table}[b]
  \caption{Summary of the limiting behaviors of $w(\omega, q_z)  $ 
  that shows relative magnitudes to the anomalous term. 
The high-$  T$ limit refers to  $ m/T \to 0 $ and $  T/\mu \gg1 $, 
while the high-$ \mu $ limit $ m/\mu \to 0 $ and $ T/\mu \ll1 $.}
  \label{table:summary}
  \vspace{0.5cm}
  \centering
  \begin{tabular}{l || c |c c c c c}
& \quad   \quad ``Massless'' limit  \quad   \quad  
& & \hspace{-4.5cm} Massive fermions
\\
&  ($ q_\para^2/m^2 \to \infty $)  & \quad   \quad   Adiabatic limit \quad  \quad  
& \quad    \quad  Diabatic processes  \quad  \quad  
\\
& & ($ q_\para^2/m^2 \to 0 $) &  (Finite $ \omega $ and/or $  q_z$)
\\
\hline \hline
 \quad  \quad  Vacuum   \quad  & 1 & 0 & Nonzero from pair creation
\\
\hline
 \quad  \quad Medium   \quad  \quad  &
  \begin{tabular}{c}1 \\ \quad  \end{tabular}
  & \begin{tabular}{c} Nonzero \\ $  [$1 in high-$T \, (\mu)  $ limit $  ]$ \end{tabular}
 & \begin{tabular}{c} Nonzero from pair creation \\ and Landau damping \end{tabular}
 \end{tabular}
\end{table}

}

In the following, we find that the AWI (\ref{eq:anomaly-LLL-1}) takes various forms 
depending on the scalar function $ \Pi_\para $ stemming from the polarization tensor. 
Such behaviors crucially depend on whether fermions are massive or massless 
and whether there is the medium contribution or not. 
We provide summary of the characteristic behaviors before going into detailed discussions. 
The reasons/interpretations leading to those results are given in the following subsections.

In the strictly massless case, it is well-known that the divergence of the axial current 
is saturated by the anomalous term proportional to the pseudoscalar inner product 
of the electromagnetic field. 
It is thus useful to take this case as a reference result to other limits. 
Inserting the polarization tensor (\ref{eq:Pi-massless}) into Eq.~(\ref{eq:anomaly-LLL-1}), 
one finds the AWI 
\begin{eqnarray}
{\rm F.T.} \ \pd_\mu j_A^\mu 
=  s_f   \rho_B  \frac{q_f}{\pi} \tilde E_\para (q) \, e^{-\frac{|\bq_\perp|^2}{2|q_fB|}} 
=  \frac{ q_f^2}{2\pi^2} \tilde \bE (q) \cdot \bB  \, e^{-\frac{|\bq_\perp|^2}{2|q_fB|}} 
\label{eq:anomaly-LLL-2}
\, .
\end{eqnarray}
Remember that $\tilde E_\para $ is the component of the electric field 
automatically projected in the magnetic-field direction in Eq.~(\ref{eq:anomaly-LLL-1}) 
and depends on $ q^2_\para$ and $ |\bq_\perp|^2 $ in general, 
while $ B $ is a constant strength. 
The same structure arises repeatedly below. 
This AWI (\ref{eq:anomaly-LLL-2}) reproduces the chiral anomaly 
in the (3+1) dimensions up to the Gaussian factor. 
We will discuss the Gaussian factor below and put it aside at this moment.

Notice that both the divergence of the axial current 
and the product of the electromagnetic fields are dimension-four quantities. 
Therefore, the remaining factor in Eq.~(\ref{eq:anomaly-LLL-2}) is just a constant 
which is sometimes referred to as the anomaly coefficient. 
The same applies to the general form of $ \Pi_\para $ in Eq.~(\ref{eq:anomaly-LLL-1}). 
Since $ \Pi_\para $ always has the overall factor of $ q_f B $ from the Landau degeneracy factor, 
the remaining part has to be only a function of dimensionless quantities. 
Including all the contributions to $ \Pi_\para $ obtained in Eq.~(\ref{eq:Pi_medium-1}), 
we find that 
\begin{subequations}
\label{eq:AWI-thermal}
\begin{eqnarray} 
&& {\rm F.T.} \ \pd_\mu j_A^\mu = w(\omega, q_z) 
 \frac{ q_f^2}{2\pi^2} \tilde \bE(q) \cdot \bB  \, e^{-\frac{|\bq_\perp|^2}{2|q_fB|}}  
\, ,
\\
&& w(\omega, q_z) := 1 - I \big( \frac{q_\parallel^2}{4m^2} \big) 
 +
 m^2  \int_{-\infty}^\infty \frac{d p_z}{ \epsilon_p}  
 \frac{    n_+(\epsilon_p)  + n_-(\epsilon_p)   }{ q_\para^2\sqrt{1- 4m^2/q_\para^2 }  } 
\Big( \frac{ \epsilon_p^+ s_+ }{ p_z - p_z^+ } - \frac{ \epsilon_p^- s_-}{ p_z - p_z^- } \Big)
  \, .
\end{eqnarray}
\end{subequations}
The first term in $  w(\omega, q_z)$ is the same anomalous term as in 
the massless case (\ref{eq:anomaly-LLL-2}), so that 
a deviation of $  w(\omega, q_z)$ from unity quantifies 
both the vacuum and medium contributions for massive fermions. 
It may be appropriate to call Eq.~(\ref{eq:AWI-thermal}) the axial Ward identity (AWI) 
rather than the anomaly equation or something of this sort, 
because the right-hand side now has the terms other than the anomalous term. 
Those additional terms are identified with matrix elements of the pseudoscalar condensate, i.e., 
$ 2im \langle \bar \psi \gam^5 \psi \rangle $. 
In Appendix~\ref{anomaly-massive-vac}, we confirm this statement 
with an explicit computation of the axial-vector correlator.

\begin{figure}[b]
     \begin{center}
              \includegraphics[width=0.9\hsize]{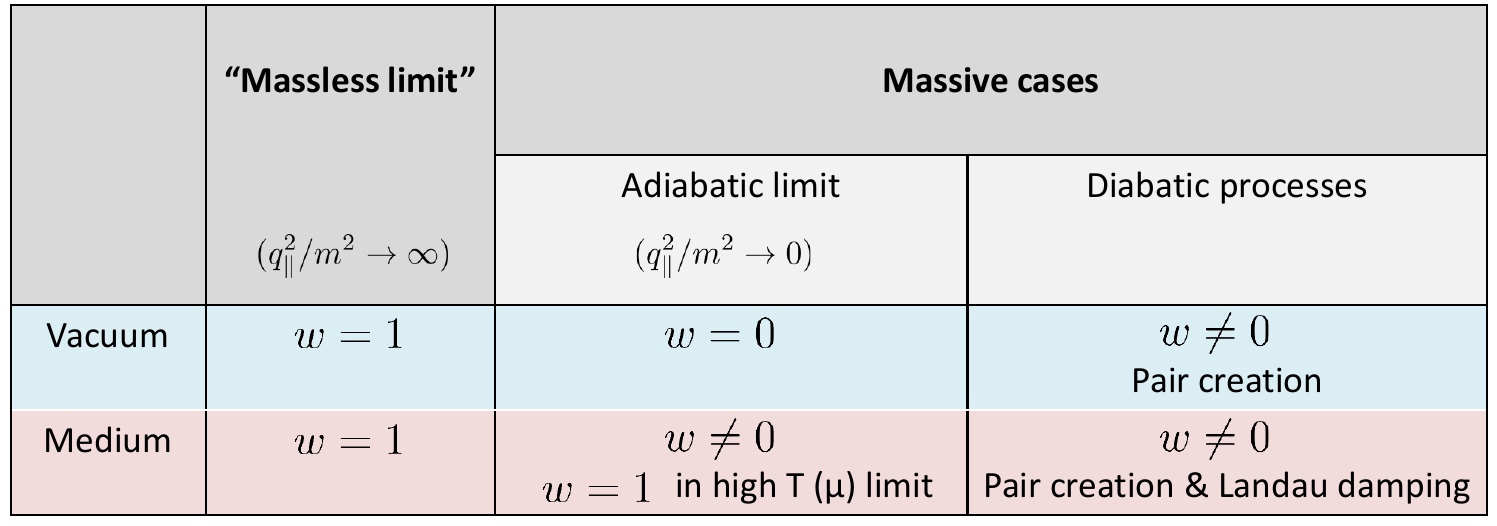}
     \end{center}
\vspace{-0.5cm}
  \caption{Summary of the limiting behaviors of $w(\omega, q_z)  $ 
  that show relative magnitudes to the anomalous term. 
The high-$  T$ limit refers to  $ m/T \to 0 $ and $  T/\mu \gg1 $, 
while the high-$ \mu $ limit $ m/\mu \to 0 $ and $ T/\mu \ll1 $.}
  \label{table:summary}
\end{figure}

We summarize the limiting behaviors defined by the dimensionless combinations 
below and in Table~\ref{table:summary}.

\begin{itemize}

\item ``Massless'' limit $q_\para^2 / m^2 \to \infty  $: 
There is no medium correction in this limit, and the result is the same 
as in vacuum with massless fermions. 
This limit needs to be defined with a finite value of $ q_\para^2 $ 
that requires a nonzero energy $ \omega $ and/or momentum $ q_z $ 
of the electric field $ \tilde E_\para $ along the magnetic-field direction. 
The absence of temperature corrections to the anomalous term was 
shown in the strictly massless case to all orders in the coupling constant \cite{Itoyama:1982up}. 

\item Adiabatic limit $q_\para^2 / m^2 \to 0  $ in vacuum: 
As explicitly shown in Sec.~\ref{sec:triangle-LLL}, this limit corresponds to constant electric fields. 
The axial current is conserved in vacuum since a massive fermion and antifermion pair 
cannot be created in an adiabatic process \cite{Ambjorn:1983hp}. 

\item Adiabatic limit $q_\para^2 / m^2 \to 0  $ in medium: 
The spectral flow occurs to the medium particles and gives rise to a nonzero divergence of the axial current. 
This is because acceleration by electric fields 
induces a momentum flip and thus a helicity flip when the spectral flow goes through 
the bottom of the parabolic dispersion relation. 
We will find that, in the high-temperature or density limit ($ m/T , \ m/ \mu \to0 $), 
the nonzero divergence takes the same form as the anomalous term in vacuum, 
though it arises from the mass-dependent part instead of the anomalous term.

\item Diabatic processes: 
A nonzero energy $ \omega $ and/or momentum $ q_z $ of the electric field $ \tilde E_\para $ 
allows for an axial-charge creation in diabatic processes only if fermions are massive. 
The relevant processes are a fermion and antifermion pair creation from a single photon 
and the Landau damping that is a scattering of a medium particle off a space-like photon. 

\end{itemize}

Overall, one can only get simple results in the adiabatic limit 
which can be understood with the spectral flow. 
However, the flow pattern depends on the fermion mass, temperature and density. 
Once diabatic processes are activated, fermions do not follow the ordered collective flow. 
Then, one needs to investigate the frequency and wavelength dependences of the AWI 
with the explicit form of the response function, which do not exhibit universal behaviors. 
The AWI takes the simple universal form in the massless case
only because those diabatic processes are kinematically prohibited.\footnote{
In a sense, this situation is similar to that in quantum Hall systems and/or topological insulators 
where clean edge currents can be measured thanks to the absence of 
non-universal metallic currents in the bulk. 
Noises are muted by inherent mechanisms. 
} 
The massless nature strongly restricts possible processes to those that 
occur in the right- and left-handed sectors independently [see comments below Eq.~(\ref{eq2})].



\subsection{Vacuum contributions}

We discuss the vacuum processes that create axial charges 
and find that massless and massive fermions behave quite differently.

\subsubsection{Massless case}

\label{sec:massless}

The AWI (\ref{eq:anomaly-LLL-2}) reproduces the chiral anomaly 
in (3+1) dimensions up to the Gaussian factor. 
The numerical factor is reproduced as a product of the Landau degeneracy factor $ \rho_B $ 
and the anomaly coefficient in the (1+1) dimensions, i.e., 
$ {\rm F.T.} \ \pd_\mu j_A^\mu  =  (q_f/\pi) E$ (see, e.g., Ref.~\cite{Peskin:1995ev}). 
This is a natural consequence of the effective dimensional reduction in the LLL. 
The total current is given by sum of the copies of the (1+1)-dimensional currents; 
Degenerate states in the momentum space correspond to 
the center positions of cyclotron motion in the coordinate space.

However, the wave function of the LLL has a finite extension in the transverse plane 
that is of the order of the cyclotron radius $ 1/ \sqrt{ | q_f B|}  $. 
The extra Gaussian factor is a manifestation of this finite extension, 
and is a function of the ratio of the cyclotron radius to the transverse resolution scale $ 1/|\bq_\perp| $. 
The Gaussian factor reduces to unity in the low-resolution limit $ |\bq_\perp|^2/|q_fB| \to 0 $. 
We discuss the relation between the AWIs from the LLL approximation 
and from the familiar triangle diagrams more explicitly in Appendix~\ref{sec:triangle-LLL}.

\begin{figure}
     \begin{center}
              \includegraphics[width=0.6\hsize]{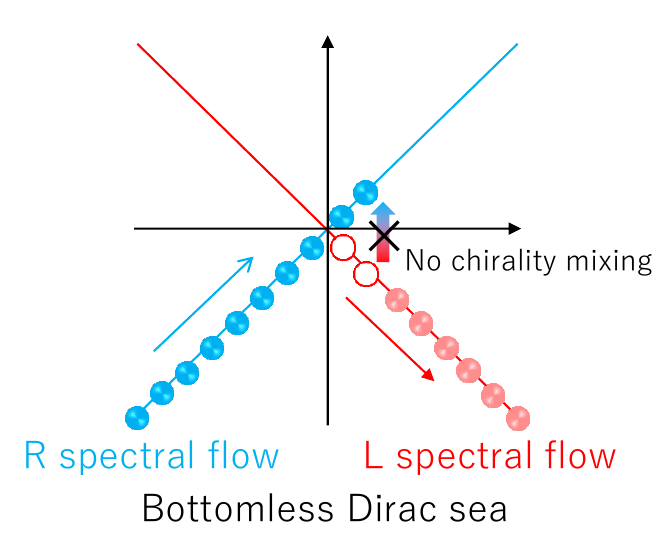}
     \end{center}
\vspace{-1cm}
\caption{Spectral flows of right- and left-handed massless fermions in vacuum.}
\label{fig:adiabatic-vacuum-massless}
\end{figure}

As well-known, the above axial-charge creation can be interpreted 
with the spectral-flow picture \cite{\cout{Blaer:1981ps,}Nielsen:1983rb, Ambjorn:1983hp} 
(see, e.g., Ref.~\cite{Peskin:1995ev, Creutz:2000bs} for pedagogical discussions). 
The spectral flow refers to an adiabatic shift of the filled states along the dispersion curve 
in response to external electric fields, 
as shown in Fig.~\ref{fig:adiabatic-vacuum-massless} for massless fermions in vacuum.

To further apply the spectral-flow picture to the massive and/or in-medium cases, 
we remind the readers of two simple but crucial points. 
The spectral flow occurs along the dispersion relation of fermions and 
it occurs only when the states on the dispersion relation are occupied. 
Note that the fermion mass enters the dispersion relation, 
while temperature and/or density enters the fermion distribution function. 
Below, we will see that those parameters control patterns of the spectral flow 
and even determine whether or not the axial-vector current is effectively conserved.

In case of the strictly massless case, the dispersion relations of the right- and left-handed fermions 
are represented with the diagonal lines in Fig.~\ref{fig:adiabatic-vacuum-massless}. 
Independent spectral flows occur along the two dispersion relations 
since there is no chirality mixing in the massless case. 
While the spectral shifts do not change the occupation number 
in the bottomless Dirac sea due to the infinity, 
a finite difference manifests itself in the infrared regime in the form of the chirality production.



\subsubsection{Massive case}

\label{sec:massive}

\begin{figure}
\begin{minipage}{0.45\hsize} 
	\begin{center} 
\includegraphics[width=\hsize]{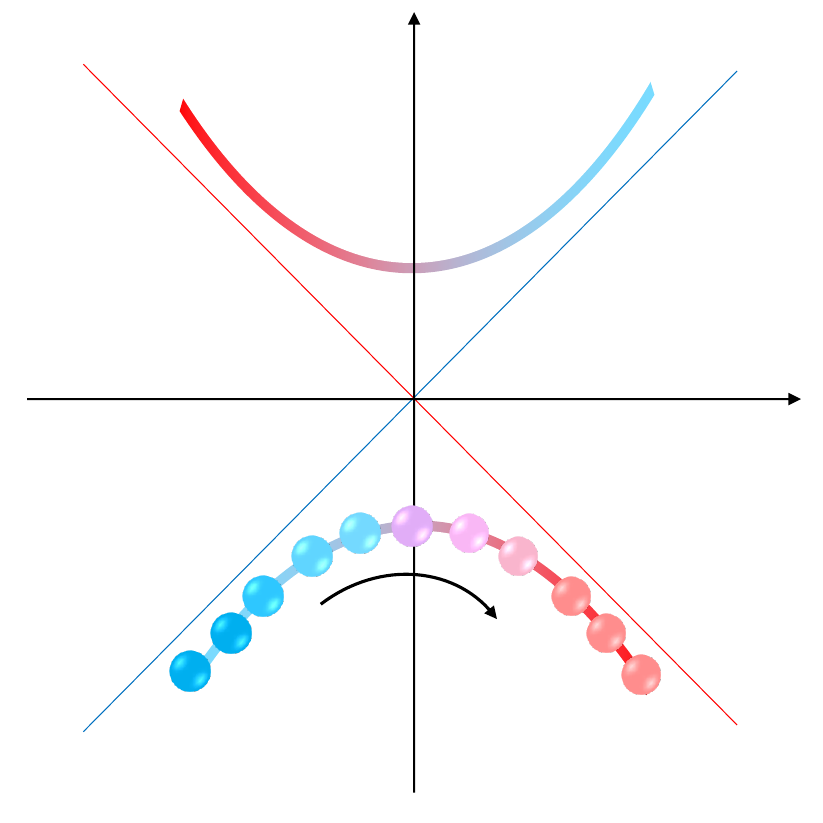}
	\end{center}
\end{minipage}
\begin{minipage}{0.45\hsize}
	\begin{center}
\includegraphics[width=\hsize]{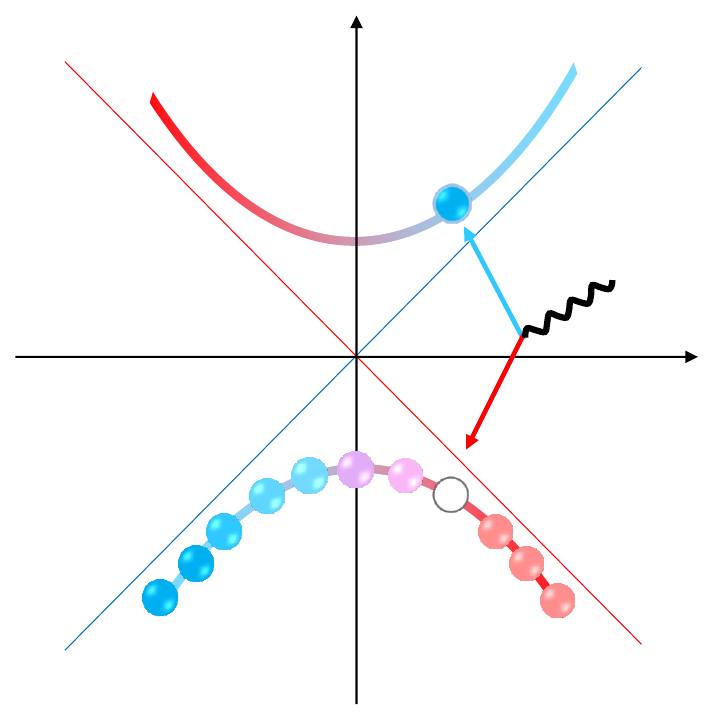}
	\end{center}
\end{minipage}
\vspace{-0.5cm}
\caption{Adiabatic spectral flow of massive fermions in vacuum (left) and 
diabatic massive pair creation from a single photon in vacuum (right).}
\label{fig:anomaly-massive-vac}
\end{figure}

Next, we show that the paths of the spectral flows are changed for massive fermions 
(cf. Fig.~\ref{fig:anomaly-massive-vac}), giving rise to a drastic change in the AWI. 
Also, we point out that chirality mixing by a finite mass term allows for 
a {\it diabatic} axial-charge creation.

The massive polarization function $ \Pi_\para $ in Eq.~(\ref{eq:vac_massive}) provides the AWI 
\begin{eqnarray}
\label{eq:AWI}
{\rm F.T.} \ \pd_\mu j_A^\mu =
 \frac{ q_f^2}{2\pi^2} \tilde \bE (q) \cdot \bB  \Big[ \, 1- I \big( \frac{q_\parallel^2}{4m^2} \big) \,  \Big] 
  e^{-\frac{|\bq_\perp|^2}{2|q_fB|}}  
  \, .
\end{eqnarray}
The unit term in the brackets is the same anomalous term as in the massless case (\ref{eq:anomaly-LLL-2}). 
As explicitly identified in Appendix~\ref{anomaly-massive-vac}, 
the other term is the matrix element of the pseudoscalar condensate 
$ 2im \langle \bar \psi \gam^5 \psi\rangle $, 
which automatically arises from computation of the massive fermion loop. 
In terms of the classical equation of motion, the origin of this term is easily identified 
with the explicit symmetry breaking by the fermion mass term in the Lagrangian. 
Nevertheless, the matrix element needs a one-loop diagram 
with insertion of the external photon field to take a nonzero value.

Notice that the anomalous term itself is independent of the mass scale 
since it originates from the ultraviolet regularization of the bottomless Dirac sea. 
In other words, the spectral flow from the bottom of the Dirac sea is always there. 
However, the magnitude of the axial-current generation, the manifestation of chiral anomaly 
in the infrared regime, depends on the relative magnitude to the pseudoscalar-condensate term 
that depends on the fermion mass. 

One can discuss the adiabatic limit such that $ q_\para^2/m^2  \to 0 $, 
where there is no invariant energy transfer from the electric field. 
This configuration corresponds to a constant electric field 
that is time-independent and homogeneous along the magnetic-field direction 
(see Appendix~\ref{sec:triangle-LLL} for more explicit explanations). 
Note that the electric field can still have an inhomogeneity in the transverse plane 
when $ |\bq_\perp| $ is finite, giving the Gaussian factor. 
In the adiabatic limit, we have $ I(0) =1 $, so that the axial current is conserved, i.e.,  
\begin{eqnarray}
\label{eq:AWI-massive-vacuum}
\lim_{ q_\para^2/m^2 \to 0 } {\rm F.T.} \ \pd_\mu j_A^\mu = 0
  \, .
\end{eqnarray}
This is a natural result because the weak electric field, perturbatively applied 
in Eq.~(\ref{eq:anomaly-LLL-1}), cannot create massive on-shell fermions 
in an {\it adiabatic} process.  
Any finite value of mass corresponds to the ``infinite mass'' limit 
in the relative sense $ q_\para^2/m^2 \to 0 $. 
This explains why the above result is independent of the fermion mass; 
Unity $ (= m^2/m^2) $ is the only one possible dimensionless combination when $ q_\para^2 \to 0 $. 
The massless limit should be taken with a finite value of $ q_\para^2 $ as $ q_\para^2/m^2 \to \infty $, 
reproducing the massless limit~(\ref{eq:anomaly-LLL-2}) with $ I(\infty) = 0 $.

As clear in the above derivation (cf. Appendix~\ref{anomaly-massive-vac} as well), 
the conservation of the axial current (\ref{eq:AWI-massive-vacuum}) 
simply means that a far-off-shell fermion loop vanishes when the fermion mass 
is much larger than any of external momenta, $ q_\para^2/m^2 \to 0 $. 
Also, one can understand the conservation of the axial current 
in terms of the spectral flow along the parabolic curve in Fig.~\ref{fig:anomaly-massive-vac} (left). 
Since the positive- and negative-energy states are separated by the mass gap, 
the adiabatic spectral flow turns back to the bottom of the Dirac sea, 
creating no axial charge \cite{Ambjorn:1983hp, Smilga:1991xa}. 
This flow pattern is in a clear contrast to the massless case in Fig.~\ref{fig:adiabatic-vacuum-massless}. 
Below, we find that the medium contribution gives rise to an additional spectral flow.

In diabatic processes, a finite axial charge can be created 
with a finite energy transfer ($ q_\para^2/m^2 \not = 0 $) from non-constant electric fields. 
Remember that a fermion and antifermion pair can be created from a single photon 
in strong magnetic fields as discussed in Paper I. 
This process is possible only for massive fermions 
because the created pair has the same helicity along the magnetic field, 
requiring a chirality mixing at the vertex 
(see Appendix~\ref{sec:LLL-fermions} for summary of the quantum numbers 
that specify the LLL fermions and antifermions). 
Once the process occurs, the created pair carries a net helicity, 
giving rise to a nonzero contribution to the AWI.\footnote{
This can be understood from kinematics in the center-of-momentum frame of the pair 
where the momenta of a pair is in the back-to-back direction. 
The spin directions are also oriented in the opposite directions 
along the magnetic field due to the large Zeeman energy. 
Thus, the pair has the same helicity. 
}

Figure~\ref{fig:anomaly-massive-vac} (right) exhibits a pair creation from a single photon. 
The chirality mixing by a finite fermion mass is expressed with a color gradient from red to blue. 
The mixing strength is maximum at $ p_z =0 $, and the fermion wave function approaches 
the chirality eigenstate in the large momentum regime $|p_z| /m \gg 1  $. 
Therefore, when a created fermion is right-ish (left-ish) in chirality, 
an antiparticle can be left-ish (right-ish), creating a finite axial charge. 
The back-to-back kinematics is only allowed for massive fermions with this chirality mixing, 
and the diabatic pair creation is prohibited in the massless case. 
A similar prohibition mechanism is known as the ``helicity suppression'' 
in the leptonic decay of charged pions \cite{Donoghue:1992dd, Zyla:2020zbs}.




\subsection{Medium contributions}

\subsubsection{Spectral flow in medium}

Next, we consider the medium contributions. 
In contrast to the vacuum case, even a soft photon can perturb the systems 
when medium fermions are populated in the positive-energy states 
above the mass gap at finite temperature and/or density. 
In this case, the axial current will not be conserved 
if scatterings with single photons induce helicity flips. 
Including the medium contribution to $ \Pi_\para $ obtained in Eq.~(\ref{eq:Pi_medium-1}), 
we find the AWI (\ref{eq:AWI-thermal}). 
Again, there is no correction to the anomalous term itself 
since the medium corrections do not change the (superficial) UV divergence. 
The medium part, proportional to the fermion mass, can be identified with 
the medium correction to the matrix element of the pseudoscalar condensate. 
Namely, including the medium correction, the pseudoscalar condensate in the LLL reads 
\begin{eqnarray}
\label{eq:PS-condensate-LLL-medium}
2im \langle \bar \psi \gam^5 \psi \rangle
\=  \frac{ q_f^2}{2\pi^2} \tilde \bE(q) \cdot \bB  \, e^{-\frac{|\bq_\perp|^2}{2|q_fB|}}  
\left[  w(\omega, q_z) - 1 \right]
\, ,
\end{eqnarray}
with $   w(\omega, q_z)$ defined in Eq.~(\ref{eq:AWI-thermal}). 
Those terms vanish in the massless limit, i.e., $  w(\omega, q_z) \to 1 $, and 
the AWI (\ref{eq:AWI-thermal}) goes back to that in the massless limit (\ref{eq:anomaly-LLL-2}).

\begin{figure}[t]
     \begin{center}
              \includegraphics[width=0.6\hsize]{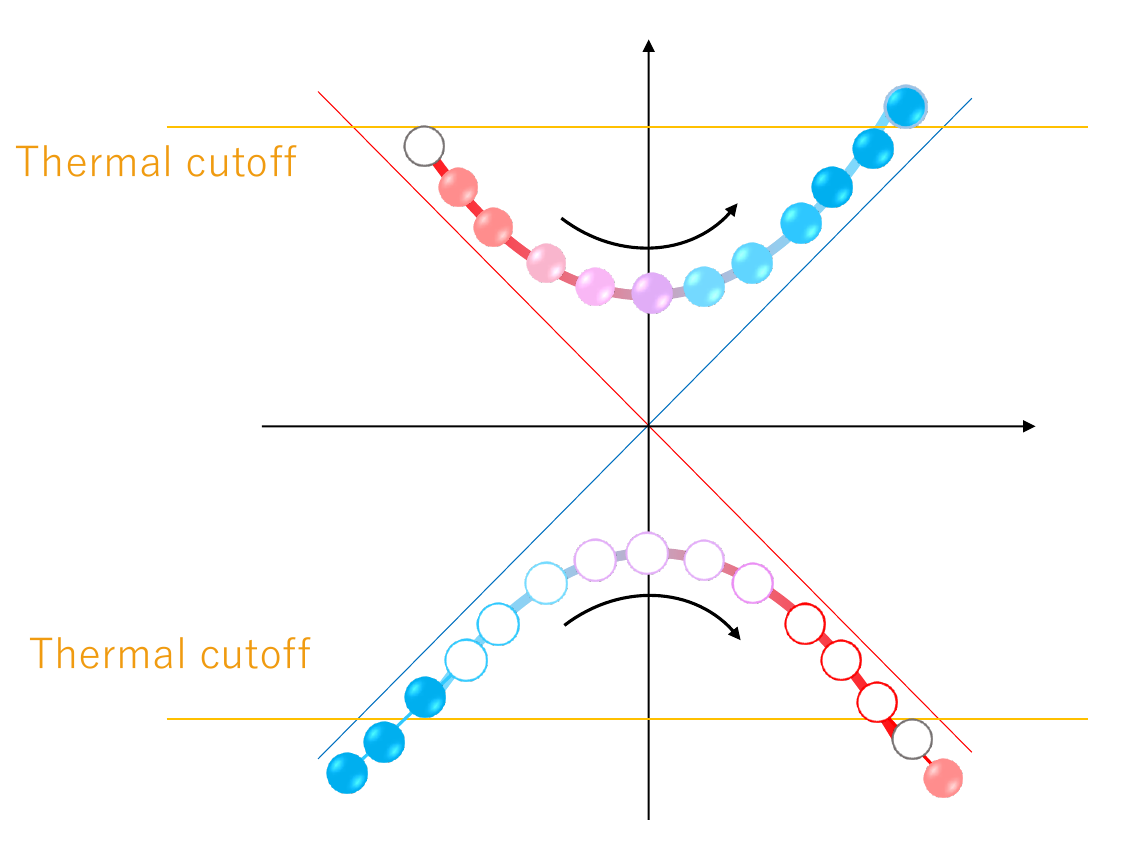}
     \end{center}
\vspace{-1cm}
\caption{Spectral flows of massive fermions and antifermions (shown as holes) in medium.}
\label{fig:adiabatic-thermal-massive}
\end{figure}

We first discuss a drastic change in the spectral flow. 
In contrast to the vacuum case where the spectral flow of massive fermions 
does not generate an axial charge [cf. Eq.~(\ref{eq:AWI-massive-vacuum})], 
the spectral flows of medium particles can create axial charges 
as illustrated in Fig.~\ref{fig:adiabatic-thermal-massive}. 
Notice that the Fermi-Dirac distribution function works as a finite cutoff to the loop momentum. 
This makes a big difference from the vacuum spectral flow beneath the {\it bottomless} Dirac sea 
(see Fig.~\ref{fig:anomaly-massive-vac} and the discussions there). 

In the presence of the thermal cutoff, 
a state just above the thermal cutoff is filled after the spectral flow on the positive-energy branch 
and a previously occupied state just below the thermal cutoff 
is removed on the other wing of the parabolic dispersion curve (cf. Fig.~\ref{fig:adiabatic-thermal-massive}). 
This process of course does not change the total particle number, but changes the helicity. 
What happens during this process is a momentum flip, and thus a helicity flip, 
across the ``valley'' of the parabolic dispersion curve. 
The same occurs in the positive-energy antiparticle branch, creating a net axial charge. 
This axial-charge creation is not anomalous in the sense that 
its origin is not related to the concept of infinity known with the ``Hilbert hotel'' \cite{gamow1988one}. 

Indeed, the above axial-charge creation originates not from the anomalous term 
but from the mass-dependent pseudoscalar condensate term. 
Remarkably, the result will turn out to have the same form as the anomalous term 
in the adiabatic and high-temperature limits (or in the adiabatic and high-density limits). 
This is simply because the axial-charge imbalance appears near the thermal cutoff 
instead of the infrared regime, so that effects of the finite curvature along the dispersion curve 
is negligible in the high-temperature or high-density limit defined as $ m/T , \ m/\mu \to 0 $.

To verify the above statements, we examine the zero frequency and momentum limits. 
Though the results turn out to be independent of the ordering of the limits, 
one can consider the following two orderings:  
\begin{subequations}
\label{eq:AWI-thermal-adiabatic}
\begin{eqnarray}
\lim_{\omega/m\to 0 } \lim_{q_z /m\to 0 } {\rm F.T.} \ \pd_\mu j_A^\mu \=
 \frac{ q_f^2}{2\pi^2} \tilde \bE(q) \cdot \bB  \, e^{-\frac{|\bq_\perp|^2}{2|q_fB|}}  
 \lim_{\omega/m\to 0 }
 2 m^2 \prj \int_{-\infty}^\infty \frac{d p_z}{\epsilon_p} 
\frac{  n_+(\epsilon_p)  + n_-(\epsilon_p)  } 
{  (2 \epsilon_p)^2  -  \omega^2 }
\, ,
\nnb
\\
\lim_{q_z/m\to 0 } \lim_{\omega /m\to 0 } {\rm F.T.} \ \pd_\mu j_A^\mu \=
 \frac{ q_f^2}{2\pi^2} \tilde \bE(q) \cdot \bB  \, e^{-\frac{|\bq_\perp|^2}{2|q_fB|}}  
 \lim_{q_z/m\to 0 }
\left(-  \frac{ m^2}{q_z} \right)  \prj  \int_{-\infty}^\infty \frac{d p_z}{\epsilon_p} 
\frac{ n_+(\epsilon_p)  + n_-(\epsilon_p) }  {  p_z - \frac{1}{2} q_z } 
\, ,
\nnb
\end{eqnarray}
\end{subequations}
where $  \prj$ denotes the principal value. 
The vacuum contribution vanishes in these limits according to Eq.~(\ref{eq:I-limits}). 
We have encountered the same integrals in Sec.~IV of Paper I 
when computing the photon masses, where the orderings do matter. 
The distribution functions can be simply replaced 
by the upper and lower boundaries of the integrals in the high-$T  $ or high-$ \mu $ limit, 
leading to simple results  
\begin{subequations}
\label{eq:high-limits}
\begin{eqnarray}
&&
\label{eq:high-limits-homogeneous}
 \lim_{\Lambda/m\to \infty }  \lim_{\omega/m\to 0 } 
2 m^2 \prj \int_{-\Lambda}^\Lambda \frac{d p_z}{ \epsilon_p}  
\frac{  1 }{  (2 \epsilon_p)^2  -  \omega^2 }
= 1
\, ,
\\
&&
\label{eq:high-limits-stationary}
 \lim_{\Lambda/m\to \infty }  \lim_{q_z/m\to 0 } 
\left(-  \frac{ m^2}{q_z} \right) \prj  \int_{-\Lambda}^\Lambda \frac{d p_z}{\epsilon_p}  
\frac{  1 } {  p_z - \frac{1}{2} q_z } 
= 1
\, ,
\end{eqnarray}
\end{subequations}
where $ \Lambda = T $ or $ \mu $. 
Note that a hierarchy $ T/\mu \gg1 $ ($ T/\mu \ll 1 $) needs to be satisfied for the above replacement, 
in addition to the high-temperature limit $m/T \to 0  $ (high-density limit $ m/\mu  \to 0$). 
Plugging the above integrals into Eq.~(\ref{eq:AWI-thermal-adiabatic}), 
one finds the AWI in the adiabatic and high-temperature (or high-density) limits 
\begin{eqnarray}
\label{eq:AWI-thermal-adiabatic-2}
\lim_{\omega/m\to 0 } \lim_{q_z /m\to 0 } {\rm F.T.} \ \pd_\mu j_A^\mu 
= \lim_{q_z/m\to 0 } \lim_{\omega /m\to 0 } {\rm F.T.} \ \pd_\mu j_A^\mu \=
 \frac{ q_f^2}{2\pi^2} \tilde \bE(q_\perp) \cdot \bB  \, e^{-\frac{|\bq_\perp|^2}{2|q_fB|}}  
\, .
\end{eqnarray}
Notice that the medium contribution to the pseudoscalar condensate 
results in the same form as the anomalous term. 
This mechanism is rather similar to the ``chiral anomaly'' in lattice models; 
The approximate linear dispersion relations are connected with each other 
as segments of the periodic dispersion relations, forming a valley similar 
to Fig.~\ref{fig:adiabatic-thermal-massive}. 
In short, we have explicitly seen that the axial charge is generated 
by the spectral flow of thermally populated fermions 
even when the spectral flow beneath the Dirac sea cannot contribute at all.

\begin{figure}
\begin{minipage}{0.45\hsize} 
	\begin{center} 
\includegraphics[width=\hsize]{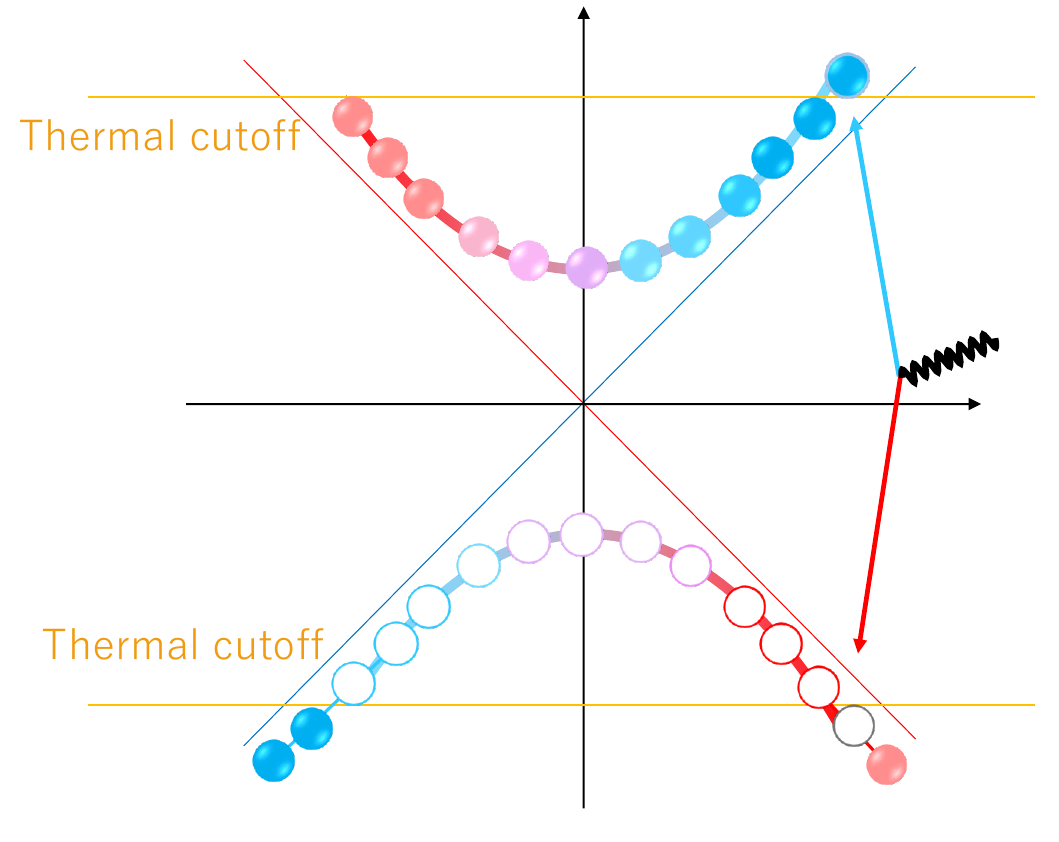}
	\end{center}
\end{minipage}
\begin{minipage}{0.45\hsize}
	\begin{center}
\includegraphics[width=\hsize]{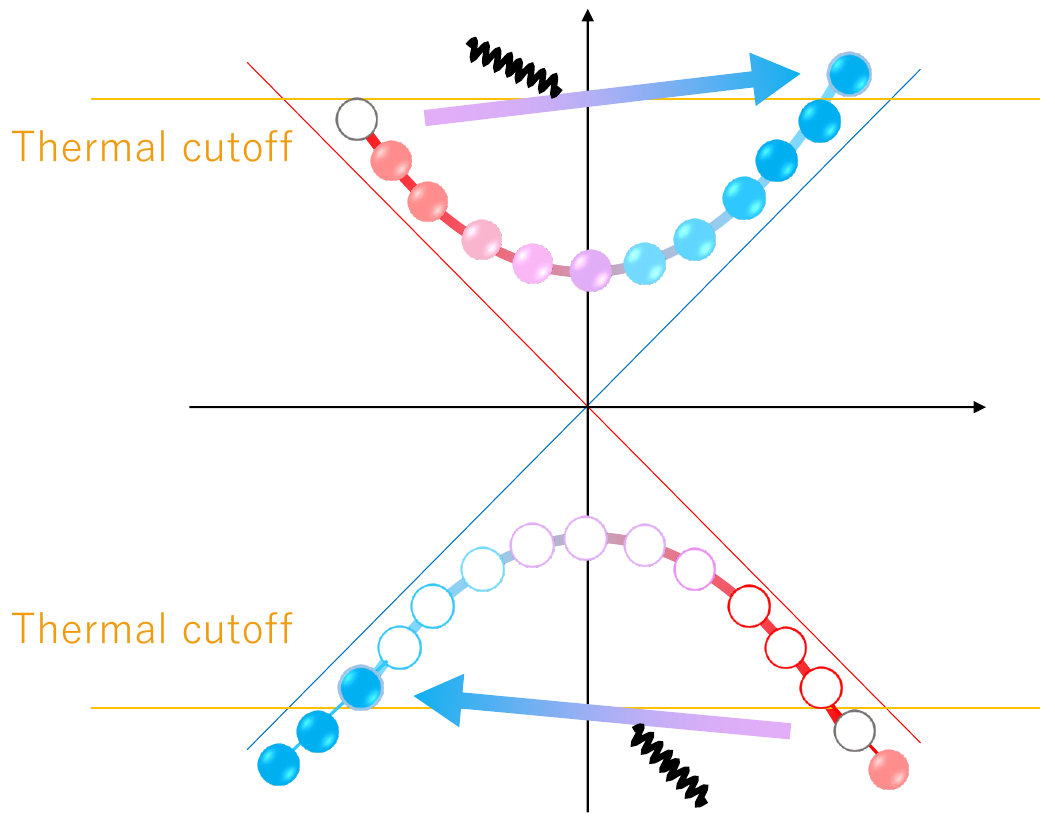}
	\end{center}
\end{minipage}
\caption{Axial-charge creation from the pair creation (left) and the Landau damping (right).}
\label{fig:diabatic-thermal-massive}
\end{figure}

\subsubsection{Diabatic processes}

When the electric field has a finite frequency $ \omega $, 
the pair creation and the Landau damping can contribute 
to the integral in Eq.~(\ref{eq:PS-condensate-LLL-medium}). 
The electric field can be both external one and of a dynamical photon 
as long as its magnitude is perturbatively weak. 
According to the optical theorem, nonzero rates of those reactions are 
directly captured by the imaginary part of the polarization tensor 
that gives a real part of the induced axial current just like 
electric currents are expressed by imaginary parts of the retarded correlators 
according to the Kubo formula.\footnote{
Note that we take the divergence of the axial current in the present case, 
which adds a complex unit and a momentum factor. 
} 
The real part of the polarization tensor also exhibits peak structures 
originating from those reactions (see Sec.~III. B in Paper I); 
The real and imaginary parts are related to each other via the dispersion integral, 
and are two sides of the same coin.

Kinematics in the relevant processes are shown in Fig.~\ref{fig:diabatic-thermal-massive}. 
We have discussed the contribution from the pair creation in Sec.~\ref{sec:massive} in vacuum case. 
The medium correction should suppress the pair creation due to the Pauli-blocking effect in the final state 
when the photon energy is smaller than the temperature and/or density scale. 
On the other hand, the Landau damping is a pure medium effect 
in which a medium fermion is scattered off an off-shell photon. 
The scatterings from one side to the other side of 
the parabolic dispersion relation is accompanied by a helicity flip. 
We have inspected those reactions in Paper I by performing 
the same integral as that in Eq.~(\ref{eq:PS-condensate-LLL-medium}).

\begin{figure}[t]
\begin{minipage}{0.49\hsize} 
	\begin{center} 
\includegraphics[width=\hsize]{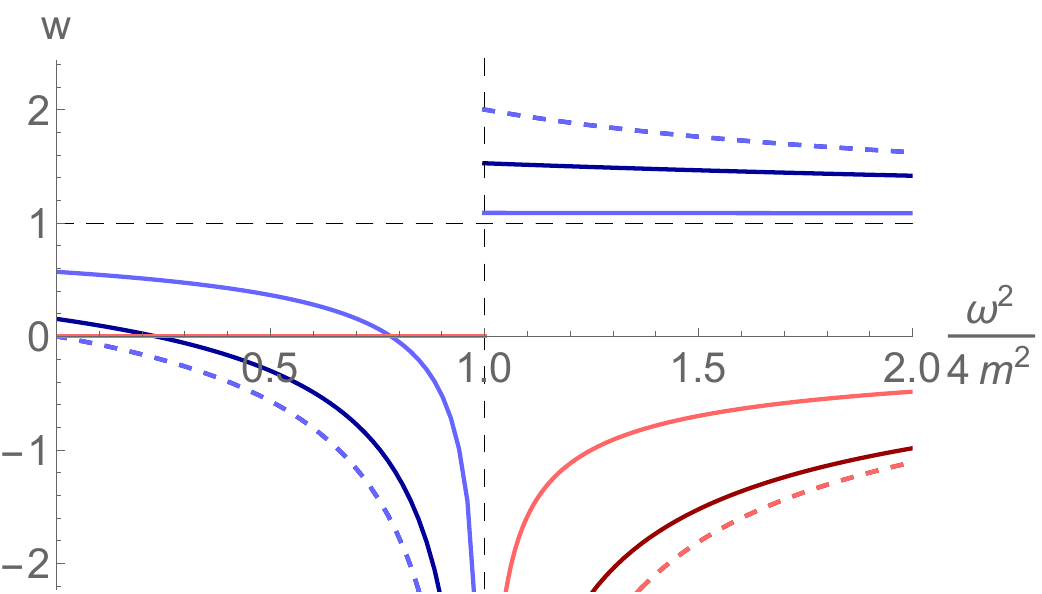}
	\end{center}
\end{minipage}
\begin{minipage}{0.49\hsize}
	\begin{center}
\includegraphics[width=\hsize]{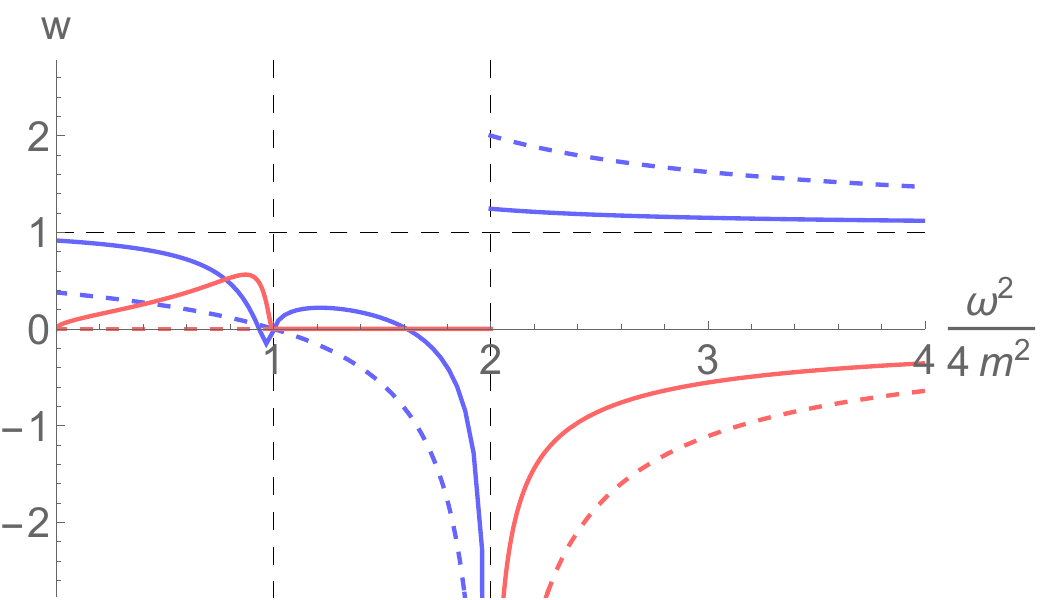}
	\end{center}
\end{minipage}
\vspace{0.5cm}
\caption{$ w(\omega, q_z) $ for the AWI at $ q_z = 0 $ 
in the homogeneous limit [left] and at $ q_z/(2m) = 1 $ [right]. 
The blue and red lines show the real and imaginary parts, respectively. 
We take temperature scales $ T/m = 0.5 $ (the darker colors) and $ T/m = 1.5 $ (the lighter colors) 
at zero density $ (\mu=0) $, while the vacuum contribution is shown with dashed lines. 
}
\label{fig:AWI-T}
\end{figure}

To understand basic behaviors of $ w (\omega, q_z) $ defined in Eq.~(\ref{eq:AWI-thermal}), 
we consider the homogeneous limit ($ q_z \to 0 $) 
\begin{eqnarray}
\label{eq:AWI-thermal-homogeneous}
 \lim_{q_z/m \to0} w (\omega, q_z) = 1 - I \big( \frac{\omega^2}{4m^2} \big) + 2 m^2 \prj \int_{-\infty}^\infty \frac{d p_z}{\epsilon_p} 
\frac{  n_+(\epsilon_p)  + n_-(\epsilon_p)  } 
{  (2 \epsilon_p)^2  -  \omega^2 }
\, ,
\end{eqnarray}
where the electric field may still have a transverse inhomogeneity $| \bq_\perp| \not = 0 $. 
The left panel in Fig.~\ref{fig:AWI-T} shows the real and imaginary parts of $ w (\omega) $ 
with blue and red lines, respectively. The colored dashed lines show the vacuum contributions. 
The adiabatic limit corresponds to the interception at $ \omega =0 $ 
that vanishes in vacuum as discussed above, but takes finite values in medium. 
We compare the low- and high-temperature cases 
with  $ T/m = 0.5 $ (darker color) and $ T/m = 1.5 $ (lighter color) at zero density $ \mu/m =0 $. 
As we increase temperature, the real part approaches 
the constant value $  w (\omega) =1$ in the massless limit. 
The imaginary part above the threshold, on the right of the dashed vertical line, 
gives the contribution from the pair creation. 
The imaginary part diverges as $ \omega $ approaches the threshold from above. 
Correspondingly, the real part also diverges as $ \omega $ approaches the threshold from below. 
Those behaviors are typical threshold behaviors in (1+1) dimensions 
and manifest themselves here because of the effective dimensional reduction in the strong magnetic fields. 
The sign of the real part depends on $ \omega $, 
which is due to a phase lag in response to time-dependent electric fields.

In the right panel of Fig.~\ref{fig:AWI-T}, 
we show the AWI at a finite value of momentum $ q_z/(2m) = 1  $. 
We take $ T/m = 1.5  $ and $\mu /m =0  $ as in the left panel. 
When $ \omega $ is small, there is a spatial region $ \omega^2 - q_z^2 <0 $ 
on the leftmost side of the dashed vertical lines. 
In this region, the imaginary part, shown with the red line, solely stems from the Landau damping. 
In the adiabatic or stationary limit ($ \omega \to 0 $), 
the imaginary part, however, vanishes for any value of $ q_z $ 
since imaginary parts of retarded correlators 
are odd functions of $ \omega $ on the general ground.
There appears a negative peak structure in the real part corresponding to 
the peak structure in the imaginary part. 
Qualitative behaviors in the other regions are common to the left panel at $ q_z =0 $. 
Both the real and imaginary parts go through the point $ w(\omega = q_z, q_z) =0 $ where $ q_\para^2=0 $.

\begin{figure}[t]
     \begin{center}
              \includegraphics[width=0.6\hsize]{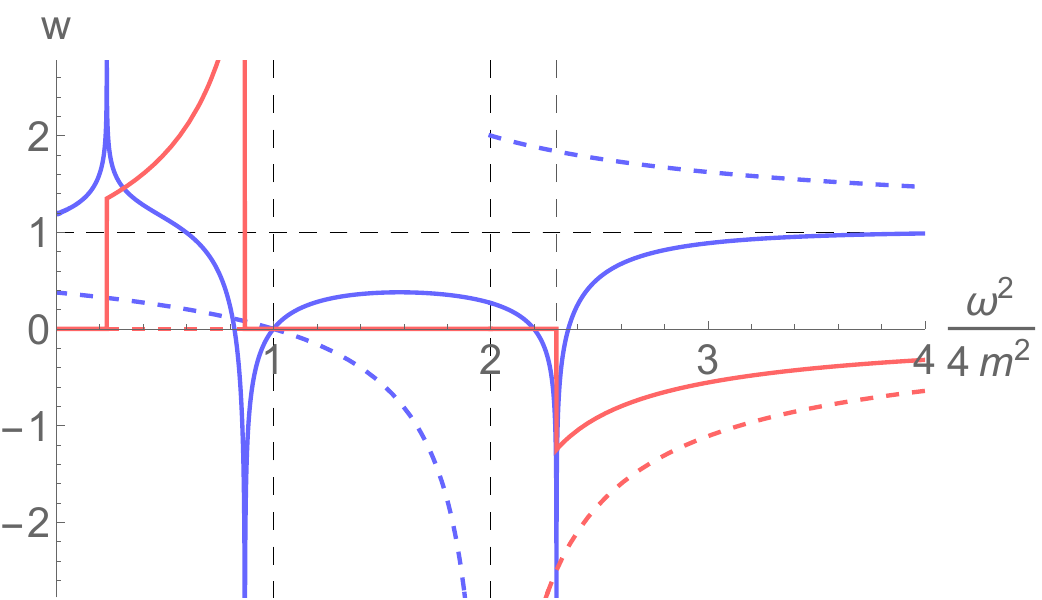}
     \end{center}
\vspace{-0.5cm}
\caption{
$ w(\omega, q_z) $ for the AWI at zero temperature and a finite chemical potential $ \mu/m = 2 $. 
We take $ q_z/(2m) = 1  $ as in Fig.~\ref{fig:AWI-T}. 
}
\label{fig:AWI-mu}
\end{figure}

Finally, we show the AWI at zero temperature and 
a finite chemical potential $ \mu/m = 2 $ in Fig.~\ref{fig:AWI-mu}. 
We take $ q_z/(2m) = 1  $ as in Fig.~\ref{fig:AWI-T}. 
Remarkably, the diverging threshold behavior at $ q_\para^2 =4m^2$ shifts to a higher energy. 
This is due to the Pauli-blocking effect; 
Since the positive-energy states are occupied up to the sharp Fermi surface, 
the pair creation can only occur in vacant states above the Fermi surface, 
requiring a larger energy of the order of the chemical potential. 
The singular behaviors at the threshold is now logarithmic 
and the threshold position splits into this one 
and a higher one that is outside the plot region (see Paper I for more discussions). 
The Landau damping also gives rise to step singularities in the imaginary part 
since it brings one occupied state below the sharp Fermi surface to 
a vacant state above the Fermi surface. 
Finite-temperature effects make vacancy below the Fermi surface, 
so that the threshold position goes back to $ q_\para^2 = 4m^2 $ at finite temperature.

%
%
%

%


\section{Summary and discussions}

In this paper, we investigated the AWI in the LLL approximation 
that is composed of the anomalous and pseudoscalar-condensate terms. 
The AWI is saturated by the anomalous term in the massless case 
irrespective of the presence or absence of the medium. 
The polarization tensor, the central quantity in this series of papers, describes 
the medium corrections to the pseudoscalar condensate 
which, therefore, should be proportional to the fermion mass. 
We showed how the AWI as a whole behaves when we vary temperature and/or density 
as well as the fermion mass and the external photon momentum. 
We classified the mechanisms of the axial-charge generation 
into the adiabatic and diabatic processes.

For the adiabatic processes, we further classified the patterns of the spectral flows 
putting an emphasis on two simple points, that is, the shape of dispersion relation and 
whether or not the states on the dispersion relation are occupied. 
The spectral flows take the paths along the dispersion relations, 
and thus there is a qualitative difference between the massless and massive cases. 
A nonzero axial charge could only stem from the spectral flow 
of the positive-energy states in the massive case 
since the anomalous term is offset by the vacuum part of the pseudoscalar condensate in the adiabatic limit. 
Therefore, a nonzero axial charge is generated by the helicity flip on the medium particles  
and does not contain a contribution from the anomalous term; 
The AWI is dominated by the medium contribution. 
It is remarkable that the medium contribution, nevertheless, agrees with the anomalous term 
in the high-temperature or high-density limit as compared to the fermion mass, 
where the curvature of the parabolic dispersion relation is negligible. 
We showed those limiting behaviors in an analytic way 
and would like to add that there are possible contributions from the higher Landau levels 
in such limits, which needs to be investigated in the future works.


We also investigated the diabatic contributions to the axial-charge generation 
which can be finite only for massive fermions. 
The relevant physical processes are the pair creation from a single photon and the Landau damping. 
Those processes are not anomalous in the sense that 
they are nothing to do with the ultraviolet divergence. 
Thus, there are possible higher-order radiative corrections. 
Such corrections in QED may be small in many of relevant applications in vacuum. 
However, when scatterings with medium particles happen so frequently in high-energy-density states, 
such corrections can be significant even in weak-coupling theories. 
In addition, there are possible contributions from the higher Landau levels. 
Remember that the AWIs from the LLL approximation and the triangle diagrams 
only agree with each other in the homogeneous limit $ |\bq_\perp|^2/|q_fB| \to 0 $ 
as explicitly shown in Appendix~\ref{sec:triangle-LLL}. 
The discrepancy needs to be compensated by contributions from the higher Landau levels, 
and can be significant when the higher Landau levels are excited 
at the temperature and/or density scale comparable to the magnetic-field strength. 
Such environments can be realized in relativistic heavy-ion collisions, neutron star, and the early universe 
depending on detailed situations. 
We leave those points and applications as future works.

Studying the axial-charge dynamics provides implications for various systems. 
The amount of the created axial charge determines the magnitude of the CME currents 
induced in relativistic heavy-ion collisions \cite{Kharzeev:2015znc, Skokov:2016yrj, 
Hattori:2016emy, STAR:2021mii} 
and the Weyl and Dirac semimetals \cite{Burkov:2015hba, Gorbar:2017lnp, 2018RvMP...90a5001A}. 
The axial-charge dynamics even plays a crucial role when the magnetic fields are dynamical 
and are subject to the chiral plasma instability that has been discussed as a possible origin of 
the strong magnetic fields in the early universe and neutron stars/magnetars \cite{Turner:1987bw, 
Carroll:1989vb, Garretson:1992vt, Joyce:1997uy, Giovannini:2003yn, Laine:2005bt, 
Akamatsu:2013pjd, Akamatsu:2014yza}. 
The time evolution of the axial-charge density should be tracked 
consistently coupled to the amplification of the magnetic fields. 
In addition to those CME-induced phenomena, one can study 
the axial-charge dynamics revealing itself in the optical activities. 
Such directions have been pursued with the Weyl and Dirac semimetals 
\cite{PhysRevB.91.035114, PhysRevB.93.075114, PhysRevB.93.085426, PhysRevB.98.075203, 
PhysRevB.100.085436, song2016detecting, rinkel2017signatures, hui2019optical, 
yuan2020discovery, PhysRevB.101.235156, PhysRevB.103.L041301, PhysRevB.103.075114, 
 PhysRevLett.120.037403, PhysRevLett.105.057401, PhysRevLett.126.247202, kargarian2015theory, 
tokman2020inverse}. 
It is interesting to investigate possible optical signals 
form the quark-gluon plasma and from the outer space as well.

There is another implication for the production of axion or any hypothetical pseudoscalar particles 
via the Primakoff effect \cite{Sikivie:1983ip, Sikivie:1985yu, Raffelt:1985nk, Raffelt:1987im, Raffelt:1990yz}. 
This mechanism can be understood as a mixing process between a single photon 
and the single pseudoscalar particle in strong magnetic fields of stars or galaxies. 
The similar idea leads to the light-shining-through-wall experiment in laboratories 
(see, e.g., Refs.~\cite{Gies:2008wv, Redondo:2010dp} for reviews). 
The two-point function between the axial and vector currents studied in this paper 
determines the mixing strength with the inclusion of the screening effect in medium. 
The screening effect can be important in stars and galaxies, giving rise to the nontrivial peak structures 
in the photon-energy and -momentum dependences shown in this paper. 
This deserves further studies in fundamental physics.

\cout{

To understand patterns of the spectral flows, one should notice two simple points. 
They are the dispersion relation of fermions and 
whether or not the states on the dispersion relation are occupied; 
The fermion mass enters the dispersion relation, 
while temperature and/or density enters the fermion distribution function. 
We show in an analytic way that those parameters control patterns of the spectral flow 
and even determine whether or not the axial-vector current is effectively conserved. 
In the high temperature or density limit over the fermion mass, the AWI has only a contribution 
that takes the same form as the anomalous term, 
but actually originates from the limiting form of the thermal contribution to the mass-dependent part. 
The spectral flow in the negative-energy states turns back to the bottomless Dirac sea 
along the parabolic dispersion relation, 
and does not leave an observable consequence in the infrared regime. 
Instead, another spectral flow occurs in the thermally occupied positive-energy states, 
inducing helicity flipping at the bottom of the parabolic valley. 
Technically, this can be seen as the cancellation between the anomalous term 
and the vacuum contribution to the mass-dependent part. 
The mass-dependent part, therefore, plays a crucial role in the adiabatic processes 
and is, furthermore, exclusively responsible for the diabatic processes.  

}

\vspace{1cm}

{\it Acknowledgments}.--- 
KH thanks Patrick Copinger, Kenji Fukushima, Yoshimasa Hidaka, 
Xu-Guang Huang, Hidetoshi Taya, and Di-Lun Yang for discussions. 
This work is supported in part by JSPS KAKENHI under grant Nos.~20K03948 and 22H02316. 

\appendix

\section{Massless fermions and antifermions in the LLL}

\label{sec:LLL-fermions}

We summarize the quantum numbers characterizing the LLL fermions and antifermions 
to clarify somewhat confusing terminologies. 
For clarity, we here assign the following terminologies.

``Chirality'' is an eigenvalue of $ \gam^5 $ that specifies the handedness of the spinor $ \psi_{R/L}  $, 
i.e., $ \gam^5 \psi_{R/L}  = \pm \psi_{R/L}  $ with the upper/lower signs for $ R/L $. 
``Axial charge'' is the difference between the number densities 
$ j_A^0 = n_R - n_L  $ created by the spinors $ \psi_{R/L} $. 
Since antiparticles are counted in the net number densities with negative signs, 
they contribute to the axial charge with opposite signs to the particle contributions. 
Those correspondences are summarized in Table~\ref{table:chiral_fermions}.

``Helicity'' is defined by the relative direction between spin and momentum as 
$ h = S_z  p_z / |p^0|   $ with the spin operator $S_z = i \gam^1 \gam^2/2 $. 
The correspondence to chirality is made clear if one refers to the Dirac operator. 
The fermion spinor $  \psi _\LLL  $ for the LLL should satisfy the following Dirac equation (see Paper I) 
\begin{eqnarray}
\label{eq:Dirac-LLL}
 (p^0 - s_f  p_z \gam^5  ) \psi _\LLL = 0
 \, ,
\end{eqnarray}
where we used the fact that the LLL spinor is an eigenfunction of the spin operator 
$2 S_z \psi _\LLL  =  (i \gam^1 \gam^2 ) \psi _\LLL=  s_f \psi _\LLL $. 
Therefore, one finds that 
\begin{eqnarray}
 h \, \psi_\LLL = \frac{1}{2} \sgn(p^0)  \gam^5 \psi_\LLL 
\, ,
\end{eqnarray}
where $ (\gam^5)^2 =1 $. 
The relative sign between the chirality and the helicity depends on the sign of the energy. 
Therefore, $\psi_\LLL^R$ ($\psi_\LLL^L$) contains the right-handed helicity particles 
and left-handed helicity antiparticles 
(the left-handed helicity particles and right-handed helicity antiparticles). 
The particles and (positive-energy) antiparticles should have opposite helicities 
because they move in the same directions but have opposite spin directions due to the Zeeman effect. 
Helicity is independent of the sign function $ s_f $, 
because both the spin and momentum directions depend on $ s_f $.\footnote{ 
Notice that the Dirac equation (\ref{eq:Dirac-LLL}) is satisfied with the dispersion relations 
$   p^0 = \pm s_f p_z $ {\it without} the symbol of absolute value. 
The upper and lower signs are for the right- and left-handed chirality eigenstates, respectively, 
and should not be mixed up with the signs for positive- and negative-energy solutions. 
The sign of $ p^0 $ depends on that of $ p_z $. 
}

Massive fermions are none of those eigenstates, 
and are the mixed states of them as shown in Figs.~\ref{fig:anomaly-massive-vac} 
and \ref{fig:adiabatic-thermal-massive} with the color gradient. 
Some scattering processes require such chirality mixing to occur with finite amplitudes.

\begin{table}[b]
  \caption{Quantum numbers that specify the massless LLL fermions and antifermions. 
  The upper and lower signs are for the right- and left-handed chirality sectors, respectively.}
  \label{table:chiral_fermions}
  \vspace{0.8cm}
  \centering
  \begin{tabular}{l || c c c c c c}
  & Spin $ (S_z) $ & Chirality $ (\gam^5) $ & Axial charge $ (j_A^0) $ 
  & Momentum $ (p_z) $ & Helicity $ (S_z p_z/|p_z|) $
\\
\hline \hline
Particles  & $+ s_f \frac12 $ & $ \pm 1 $ & $ \pm 1$ & $ \pm s_f  $ & $ \pm \frac12 $
\\
\hline
Antiparticles  & $ - s_f \frac12 $ & $ \pm 1 $ & $ \mp 1$ & $ \pm s_f  $ & $ \mp \frac12 $
  \end{tabular}
\end{table}

\section{Chiral anomaly diagram with massive LLL fermions}

\label{anomaly-massive-vac}

In this Appendix, we directly compute the anomaly diagram with massive fermions in the LLL 
and demonstrate that it splits into the anomalous term 
and the matrix element of the pseudoscalar condensate. 
One can also explicitly confirm that the anomalous term is independent of the fermion mass, 
while the latter is proportional to the fermion mass required by the chirality mixing 
in the pseudoscalar fermion bilinear.

We examine the two-point function composed of the LLL fermion loop:  
\begin{eqnarray}
\label{eq:axial-correlator}
\int d^4 x \, e^{-i q x} \langle k | \pd_\mu j_A^\mu(x) | 0 \rangle
= \rho_B e^{- \frac{ |\bq_\perp|^2}{2|q_f B|} } 
(2\pi)^2  \delta^{(2)}( k_\para - q_\para) i q_\mu \Pi_{1+1 (A)}^{\mu\nu} (q_\parallel) \epsilon^{\ast}_\nu (q) 
\, .
\end{eqnarray}
The Landau degeneracy factor and the Gaussian come from 
the transverse-momentum integral as in the computation of the vector-current correlator (cf. Paper I). 
The residual (1+1)-dimensional longitudinal part is given as 
\begin{eqnarray}
iq_\mu \Pi_{1+1 (A)}^{\mu\nu} (q_\parallel) 
= (-1) ( - i q_f )\int \!\! \frac{d^2p_\parallel}{(2\pi)^2} 
\frac{\tr [\, \sla q_\para \gam^5 i (\sla p_\parallel + m) \prj_+
\gam_\parallel^\nu i ( \, (\sla p_\parallel + \sla q_\parallel) + m\, ) \prj_+ \, ]}
{ (p_\parallel^2 -m^2)( \,  (p_\parallel + q_\parallel)^2 -m^2 \,)}
\, .
\end{eqnarray} 
The two-dimensional integral has a (superficial) logarithmic divergence, 
and we use the dimensional regularization.

It is important to make sure the properties of the $ \gam^5 $ to apply the dimensional regularization. 
The $ \gam^5 $ is extended in such a way that 
$ \{\gam^5, \gam^\mu\} =0 $ for $ \mu = 0,1,2,3 $ 
and $ [ \gam^5, \gam^\mu ] =0 $ for the other extra components. 
The extra components in the momentum is denoted with tilde like $ \tilde  \ell ^\mu$, 
which corersponds to $\ell_\perp ^\mu $ in the notations of Ref.~\cite{Peskin:1995ev}.\footnote{ 
We do not use their notations with $ \para,\perp $ 
to avoid possible confusions with the parallel and perpendicular components 
with respect to the magnetic field in the present paper.} 
The external momenta $ q^\mu,  k^\mu $ have vanishing components 
in the extra dimensions, $ \tilde q^\mu =\tilde  k^\mu =0$. 
The extra components of $ \gam ^\mu$ commute with $ \gam^5 $, i.e., $ [\tilde {\sla \ell}, \gam^5] = 0 $. 
Then, one can show an identity 
\begin{eqnarray}
\label{eq:id-gam5}
 \sla q_\para\gam^5 = \gam^5  ( \sla p_\para - m)  + ( \sla q_\para + \sla p_\para - m )  \gam^5 
+ 2 ( m - \tilde {\sla p}_\para )  \gam^5
\, .
\end{eqnarray}

Plugging the identity (\ref{eq:id-gam5}) into the (1+1)-dimensional part, we have 
\begin{eqnarray} 
i q_\mu \Pi_{1+1 (A)}^{\mu\nu} (q_\parallel) \epsilon^{\ast}_\nu (q) 
\= q_f \epsilon^{\ast}_\nu (q)   \int \!\! \frac{d^d p_\parallel}{(2\pi)^d} 
\left[\,
\frac{\tr [\,   \gam^5  \prj_+ \gam_\parallel^\nu ( \, (\sla p_\parallel + \sla q_\parallel) + m\, )   \, ]}
{    (p_\parallel + q_\parallel)^2 -m^2  }
 + \frac{\tr [\,  \gam^5 \prj_+ (\sla p_\parallel + m)  \gam_\parallel^\nu  \, ]}{ p_\parallel^2 -m^2 }
\, \right]
\nnb 
&&
- 2 q_f  \epsilon^{\ast}_\nu (q)  \int \!\! \frac{d^d p_\parallel}{(2\pi)^d} 
\frac{\tr [\,  \tilde {\sla p}_\para \gam^5 (\sla p_\parallel + m) \prj_+
\gam_\parallel^\nu  ( \, (\sla p_\parallel + \sla q_\parallel) + m\, ) \prj_+ \, ]}
{ (p_\parallel^2 -m^2)( \,  (p_\parallel + q_\parallel)^2 -m^2 \,)}
\nnb
&& + 2im \langle \bar \psi \gam^5 \psi \rangle
\label{eq:Pi-A-2}
\, .
\end{eqnarray}
Each of the two terms in the square brackets vanishes 
because the integral and trace vanish if one picks up 
the momentum and mass factors in the trace, respectively. 
The last term is identified with the matrix element of the pseudoscalar condensate 
\begin{eqnarray}
 \langle \bar \psi \gam^5 \psi \rangle
 =   ( -1)  [ - i q_f  \epsilon^{\ast}_\nu (q)  ]    \int \!\! \frac{d^2p_\parallel}{(2\pi)^2} 
\frac{\tr [\,  \gam^5 i (\sla p_\parallel + m) \prj_+
  \gam^\nu_\para i ( \, (\sla p_\parallel + \sla q_\parallel) + m\, ) \prj_+ \, ] }
{ (p_\parallel^2 -m^2)( \,  (p_\parallel + q_\parallel)^2 -m^2 \,)}
\, .
\end{eqnarray}
This is the leading nonvanishing contribution from the diagram 
with one-insertion of the external photon leg. 
The remaining term in Eq.~(\ref{eq:Pi-A-2}) will result in the anomalous term.

We shall first examine the anomalous term. 
Introducing the Feynman parameter, one can make a complete square in the denominator 
$  x(p_\parallel^2 -m^2) + (1-x)( \,  (p_\parallel + q_\parallel)^2 -m^2 )
= \{ p_\parallel + (1-x)  q_\para\}^2  - \Delta_\para$ with $ \Delta_\para = m^2 - x (1-x)q_\para^2 $. 
We then shift the integral variable as $p_\para' = p_\parallel + (1-x)  q_\para  $ 
that is allowed after the divergence is regularized. 
To have a nonvanishing trace, one should pick up $ \gam^1\gam^2 $ in $ \prj_+ $. 
To have a nonvanishing integral for the extra dimension, one needs to pick up 
one $ \sla p_\para $ from the propagator. 
Then, one needs to take the photon momentum $ \sla q_\para $ in the other fermion propagator 
to make both the trace and integral nonvanishing. 
Therefore, assuming the presence of the integral, we have 
\begin{eqnarray}
\tr [\,  \tilde{ \sla p}_\para \gam^5  (\sla p_\parallel - (1-x)  \sla q_\para  + m) \prj_+
\gam_\parallel^\nu  ( \, (\sla p_\parallel + x \sla q_\parallel) + m\, ) \prj_+ \, ]
= - 2 s_f q_{\para\mu}  \epsilon^{\mu\nu12}  \tilde  p_\para^{2} 
\, .
\end{eqnarray}
The mass dependence as well as the Feynman parameter go away from the trace. 
Then, the momentum integral reads 
\begin{eqnarray}
  \int \!\! \frac{d^d p_\parallel}{(2\pi)^d}  \frac{  \tilde p_\para^{2}  } { ( p_\para^2 - \Delta_\para )^2 }
= \frac{d-2}{d} \times  \frac{(-1)i}{(4\pi)^{d/2}} \frac{d}{2}
 \frac{ \Gamma(- \frac{d-2}{2}) } {\Gamma(2)} \Delta_\para^{ \frac{d-2}{2}}
 =   \frac{ i}{ 4\pi } 
 \, ,
\end{eqnarray}
where we used $ (d-2) \Gamma(- \frac{d-2}{2}) = - 2 +  \order( d-2) $. 
The integral is finite because of the overall dimensional factor $(d-2)  $. 
All the subleading terms vanish thanks to the same overall factor, 
and thus the integral is given by the above universal form. 
Then, we obtain the (1+1) dimensional AWI 
\begin{eqnarray}
\label{eq:axial-Ward-app}
i q_\mu \Pi_{1+1 (A)}^{\mu\nu} (q_\parallel)  \epsilon^{\ast}_\nu (k) 
= - \frac{ q_f}{\pi}   \langle k | \tilde E_\para | 0 \rangle   +  2im \langle \bar \psi \gam^5 \psi \rangle
\, ,
\end{eqnarray}
where the momentum conservation in Eq.~(\ref{eq:axial-correlator}) is understood 
(with the delta function suppressed). 
As in Eq.~(\ref{eq:Ward-ids}), the Fourier spectrum of the electric field is defined as 
$\tilde E_\para (q) = - i   \epsilon^{\mu\nu}_\para q_\mu  A_\nu 
$ with $ \epsilon^{\mu\nu}_\para = \epsilon^{\mu\nu12} $.

Next, we compute the pseudoscalar condensate 
\begin{eqnarray}
 \langle \bar \psi \gam^5 \psi \rangle
=  \frac{ s_f q_f }{2} \epsilon^{\ast}_\nu (q) \int_0^1 dx
 \int \!\! \frac{d^2p_\parallel}{(2\pi)^2} 
\frac{\tr [\,  \gam^5 \gam^1 \gam^2   (\sla p_\parallel  - (1-x) \sla q_\para+ m)
\gam^\nu_\para ( \, (\sla p_\parallel + x \sla q_\parallel) + m\, )  \, ] }
{ ( p_\para - \Delta_\para)^2 }
\, .
\nnb
\end{eqnarray}
The integral is finite. 
One needs to take $ \gam^1 \gam^2 $ in $ \prj_+ $
because four gamma matrices are required for a nonzero result of the trace with one $ \gam^5 $. 
For the same reason, one needs to pick up one mass term, 
reflecting the fact that the chirality needs to be flipped. 
Then, there is no room for $ \sla p_\para $ to be included so that the integral becomes nonvanishing. 
Therefore, the trace under the momentum integral reads 
\begin{eqnarray} 
\tr [\,  \gam^5 \gam^1 \gam^2   (\sla p_\parallel  - (1-x) \sla q_\para+ m)
\gam^\nu_\para ( \, (\sla p_\parallel + x \sla q_\parallel) + m\, )  \, ] 
= m \tr [\,  \gam^5 \gam^1 \gam^2  \gam^\nu_\para \sla q_\para  \, ]
\, . 
\end{eqnarray}
The matrix element of the pseudoscalar condensate is proportional to the fermion mass 
because the pseudoscalar fermion bilinear mixes the chirality, 
requiring that the mass term induce chirality mixing odd times on the loop diagram.

The $ x $-dependent terms cancel away. 
The remaining elementary integrals are performed as 
\begin{eqnarray}
 \int_0^1 dx    \int \!\! \frac{d^2p_\parallel}{(2\pi)^2}  \frac{  1 } { ( p_\para - \Delta_\para)^2 }
= \frac{i}{4\pi}   \int_0^1 \frac{ dx }{ m^2 - x (1-x)q_\para^2}   
=  \frac{i}{4\pi m^2}   I \left( \frac{q_\para^2}{4m^2} \right)
\, ,
\end{eqnarray}
yielding the $ I $ function defined in Eq.~(\ref{eq:I}). 
Thus, the matrix element of the pseudoscalar condensate is given as 
\begin{eqnarray}
\label{eq:PS-condensate-LLL}
2im \langle \bar \psi \gam^5 \psi \rangle
=  \frac{s_f  q_f  }{\pi } \tilde E_\para (q)  I \Big( \frac{q_\para^2}{4m^2} \Big)
\, .
\end{eqnarray}
Combining the anomalous term and pseudoscalar condensate in Eq.~(\ref{eq:axial-Ward-app}) reproduces 
the AWI (\ref{eq:AWI}) obtained from the polarization function $ \Pi_\para $ in Eq.~(\ref{eq:vac_massive}). 
Explicitly, we have obtained 
\begin{eqnarray}
\label{eq:AWI-LLL-final}
{\rm F.T.} \ \pd_\mu j_A^\mu  =    \frac{q_f^2}{2\pi^2} \tilde \bE (q) \cdot \bB
 \Big[ \, 1 -   I \Big( \frac{q_\para^2}{4m^2} \Big) \, \Big]    e^{ - \frac{|\bq_\perp|^2}{2|q_fB|}} 
   \, .
\end{eqnarray}
We emphasize that both the mass-independent anomalous term 
and the mass-dependent pseudoscalar condensate originate from 
the same diagram composed of the massive fermion loop. 
The massive fermion loop vanishes as a whole in the large-mass limit, $\lim_{x\to0}  I(x) =1$. 
This clearly explains the vanishing behavior observed below Eq.~(\ref{eq:AWI}).

\section{Relation to the triangle diagrams}

\label{sec:triangle-LLL}

Finally, we would like to establish an explicit connection 
between the AWI (\ref{eq:AWI}) from the LLL approximation 
and the perturbative one from the familiar triangle diagrams (cf. Fig~\ref{fig:diagrams}). 
The triangle diagrams are perturbatively computed in the linear orders 
in the electric- and magnetic-field strengths 
and contains the ABJ anomaly in (3+1) dimensions \cite{Bell:1969ts,  Adler:1969gk}. 
On the other hand, the LLL approximation only appears after the resummation of 
one-loop diagrams to all orders with respect to insertion of the external legs for the constant magnetic field; 
The all-order resummation gives the fermion propagator the pole positions at the Landau levels. 
Thus, the triangle diagrams are contained in the resummed series 
but does not exactly correspond to the one-loop polarization tensor in the LLL. 
It is a non-trivial issue how the LLL approximation 
and the triangle diagrams are connected with each other, 
though the ABJ anomaly in (3+1) dimensions is often explained as a product of 
the Landau degeneracy factor and the (1+1) dimensional anomaly in the LLL \cite{Nielsen:1983rb}. 
In this appendix, we explicitly discuss their connection including the pseudoscalar-condensate term, 
focusing on zero temperature and density case.

Computing the massive triangle diagrams in Fig.~\ref{fig:diagrams}, one finds the AWI \cite{Adler:1969gk}
\begin{eqnarray}
\label{eq:anomaly-massive}
{\rm F.T.} \ \langle p, k | \pd_\mu j^\mu_A | 0\rangle  
\=   \frac{q_f^2}{2\pi^2} 
\cdot \frac12 [ \tilde \bE(k)  \cdot \tilde \bB (p) + \tilde \bB(k)  \cdot \tilde \bE (p) ]
\\
&& \times
\Big[ \, 1 + 
\int_0^1dx \int_0^{1-x} dy   \frac{ 2m^2 }{  x (1-x) k^2  +  y (1-y)  p^2 + 2 x y  \, p\cdot k - m^2}
\, \Big] 
\nn
\, ,
\end{eqnarray}
where $ p $ and $ k $ are the two photon momenta and 
are related to that of the axial current as $ q^\mu  = p^\mu + k^\mu $ 
according to the momentum conservation. 
The first and second terms in the brackets are the anomalous term 
and the matrix element of the pseudoscalar condensate, respectively. 
The tilde on the field strengths stands for the Fourier spectrum throughout the paper.

We show that the AWIs in Eqs.~(\ref{eq:AWI}) and (\ref{eq:anomaly-massive}) 
exactly agree with each other in a certain limit. 
This agreement occurs in the limit where 
one of the photon momentum is zero while the other is finite. 
We shall take $ p^\mu \to 0 $, and then $ q^\mu = k^\mu $ led by the momentum conservation. 
Then, the AWI (\ref{eq:anomaly-massive}) reads 
\begin{eqnarray}
\label{eq:anomaly-massive-constant}
\left. {\rm F.T.} \ \langle p, k | \pd_\mu j^\mu_A | 0\rangle \right|_{p=0} 
\=   \frac{q_f^2}{2\pi^2}
 \cdot \frac12 [ \tilde \bE(q)  \cdot  \tilde \bB(0)  +  \tilde \bB(q)  \cdot  \tilde \bE(0) ]
\Big[ \, 1 -  I \Big( \frac{q^2}{4m^2} \Big)   \, \Big]  
\, .
\end{eqnarray}
Remarkably, the integral in Eq.~(\ref{eq:anomaly-massive}) 
has reduced to the $ I $ function (\ref{eq:I}) involved in the LLL approximation. 
One can show this reduction as follows. 
We have 
\begin{eqnarray}
  \int_0^1dx \int_0^{1-x} dy   \frac{ 2m^2 }{  x (1-x) k^2  - m^2}
=  \int_0^1dx   \frac{ 2m^2 (1-x) }{  x (1-x) k^2  - m^2}
\, ,
\end{eqnarray}
and notice the identical transformations 
\begin{eqnarray}
\int_0^1dx   \frac{ 2m^2 (1-x) }{  x (1-x) k^2  - m^2} 
\= \int_0^1dx  \frac{ 2m^2 x }{  (1-x) x k^2  - m^2}
= \frac12 \int_0^1dx  \frac{ 2m^2 }{  x (1-x) k^2  - m^2}
\nnb
\= - I( k^2/4m^2)
\, .
\end{eqnarray}
The sum of the first two expressions results in the third expression that is nothing but the $ I $ function. 
When one sends the other momentum $ k^\mu $ to zero and maintains a finite value of $ p^\mu $, 
the integral reduces to the $ I $ function in the same manner. 
We consider the case of $\tilde \bE (0) =0 $. 
Summing the two contributions from $\tilde \bB (p = 0) $ and $\tilde \bB (k=0)  $ 
in Eq.~(\ref{eq:anomaly-massive}), we obtain 
\begin{eqnarray}
\label{eq:anomaly-massive-constant-final}
{\rm F.T.} \ \langle q | \pd_\mu j^\mu_A | 0\rangle 
\=   \frac{q_f^2}{2\pi^2} \tilde \bE(q)  \cdot  \tilde \bB(0)  
\Big[ \, 1 -  I \Big( \frac{q^2}{4m^2} \Big)   \, \Big]  
\, .
\end{eqnarray}
Therefore, the AWI (\ref{eq:AWI}) in the LLL agrees with 
Eq.~(\ref{eq:anomaly-massive-constant-final}) when one further assumes 
a homogeneity, i.e., $|\bq_\perp| =0 $, in the plane transverse to the magnetic field.
In this limit, the Gaussian factor goes away in Eq.~(\ref{eq:AWI}) 
and the squared momentum $  q^2$ reduces to $ q_\para^2 $.

Now, we show that the vanishing momentum limit corresponds to 
a constant electromagnetic field (though the gauge field itself depends on the spacetime coordinate). 
Remember that the LLL contribution is computed in the homogeneous limit. 
For simplicity, we may consider a parallel electromagnetic field in the third spatial direction. 
Such a parallel field can be described by a linear gauge configuration 
\begin{eqnarray}
A^\mu (x) = ( - c_0 z E_0 (\bx_\perp) , - d_1 y  B_0(t,z) ,  d_2 x  B_0(t,z)  , -c_3 t E_0 (\bx_\perp) )
\, ,
\end{eqnarray}
with constants such that $ c_0 + c_3 =1 $ and $ d_1 + d_2 =1 $. 
One can still let the fields have spacetime dependences within $  E_0 (\bx_\perp) $ and $  B_0(t,z) $. 
The gauge transformation does not change the following observations, 
because we examine the strengths of the electric and magnetic fields 
\begin{subequations}
\begin{eqnarray}
\tilde E_z(p) \=  \tilde E_0(\bp_\perp) 
[c_3 \omega \delta'( \omega)  \delta(p_z) + c_0 p_z \delta'(p_z)  \delta(\omega)]
=  (2\pi)^2 \delta(\omega) \delta(p_z) \tilde E_0 (\bp_\perp)
\, ,
\\
\tilde B_z(p) \=  \tilde B_0(\omega,p_z) 
[d_2 p_x \delta'( p_x)  \delta(p_y) + d_1 p_y  \delta(p_x) \delta'(p_y)  ]
=  (2\pi)^2  \delta(p_x) \delta(p_y) \tilde B_0 (\omega,p_z)
\, ,
\end{eqnarray}
\end{subequations}
where $ \omega = p^0 $. 
The prime on the delta function stands for the derivative with respect to the argument. 
One can read off the Fourier spectra. 
When the electric field is independent of $ t $ and $  z$, 
its Fourier spectrum corresponds to vanishing $ \omega $ and $ q_z $ 
in spite of the facts that $ q^\mu $ is the momentum of the gauge field 
and the gauge field itself depends on $ t $ and $  z$. 
In a similar manner, when the magnetic field is independent of $ x $ and $  y$, 
its Fourier spectrum corresponds to vanishing $ q_x $ and $ q_y $. 
When the electromagnetic field is independent of all the spacetime coordinates, 
one gets the four-dimensional delta functions  
\begin{subequations}
\begin{eqnarray}
\tilde E_z(p) \=   (2\pi)^4   \delta^{(4)}(p) E
\, ,
\\
\tilde B_z(p) \=   (2\pi)^4 \delta^{(4)}(p) B
\, ,
\end{eqnarray}
\end{subequations}
with $ E $ and $B  $ being constant strengths. 
Therefore, the AWIs in Eqs.~(\ref{eq:AWI}) and (\ref{eq:anomaly-massive}) 
exactly agree with one another in the limits of constant magnetic field 
and of the homogeneous electric field over the extension of cyclotron orbits. 
As already discussed in Sec.~\ref{sec:massless}, 
the absence of the Gaussian factor can be understood intuitively.


It is interesting to extend the above relation to finite temperature and/or density. 
The AWI in the LLL approximation is already given in Eq.~(\ref{eq:AWI-thermal}). 
The massive triangle diagrams may be straightforwardly extended to 
finite temperature and/or density as well. 
We leave explicit computations as an open issue.

\bibliography{bib}

\begin{thebibliography}{73}%
\makeatletter
\providecommand \@ifxundefined [1]{%
 \@ifx{#1\undefined}
}%
\providecommand \@ifnum [1]{%
 \ifnum #1\expandafter \@firstoftwo
 \else \expandafter \@secondoftwo
 \fi
}%
\providecommand \@ifx [1]{%
 \ifx #1\expandafter \@firstoftwo
 \else \expandafter \@secondoftwo
 \fi
}%
\providecommand \natexlab [1]{#1}%
\providecommand \enquote  [1]{``#1''}%
\providecommand \bibnamefont  [1]{#1}%
\providecommand \bibfnamefont [1]{#1}%
\providecommand \citenamefont [1]{#1}%
\providecommand \href@noop [0]{\@secondoftwo}%
\providecommand \href [0]{\begingroup \@sanitize@url \@href}%
\providecommand \@href[1]{\@@startlink{#1}\@@href}%
\providecommand \@@href[1]{\endgroup#1\@@endlink}%
\providecommand \@sanitize@url [0]{\catcode `\\12\catcode `\$12\catcode
  `\&12\catcode `\#12\catcode `\^12\catcode `\_12\catcode `\%12\relax}%
\providecommand \@@startlink[1]{}%
\providecommand \@@endlink[0]{}%
\providecommand \url  [0]{\begingroup\@sanitize@url \@url }%
\providecommand \@url [1]{\endgroup\@href {#1}{\urlprefix }}%
\providecommand \urlprefix  [0]{URL }%
\providecommand \Eprint [0]{\href }%
\providecommand \doibase [0]{http://dx.doi.org/}%
\providecommand \selectlanguage [0]{\@gobble}%
\providecommand \bibinfo  [0]{\@secondoftwo}%
\providecommand \bibfield  [0]{\@secondoftwo}%
\providecommand \translation [1]{[#1]}%
\providecommand \BibitemOpen [0]{}%
\providecommand \bibitemStop [0]{}%
\providecommand \bibitemNoStop [0]{.\EOS\space}%
\providecommand \EOS [0]{\spacefactor3000\relax}%
\providecommand \BibitemShut  [1]{\csname bibitem#1\endcsname}%
\let\auto@bib@innerbib\@empty
\bibitem [{\citenamefont {Hattori}\ and\ \citenamefont
  {Itakura}(2022)}]{Hattori:2022uzp}%
  \BibitemOpen
  \bibfield  {author} {\bibinfo {author} {\bibfnamefont {Koichi}\ \bibnamefont
  {Hattori}}\ and\ \bibinfo {author} {\bibfnamefont {Kazunori}\ \bibnamefont
  {Itakura}},\ }\bibfield  {title} {\enquote {\bibinfo {title} {{In-medium
  polarization tensor in strong magnetic fields (I): Magneto-birefringence at
  finite temperature and density}},}\ }\href@noop {} {\  (\bibinfo {year}
  {2022})},\ \Eprint {http://arxiv.org/abs/2205.04312} {arXiv:2205.04312
  [hep-ph]} \BibitemShut {NoStop}%
\bibitem [{\citenamefont {Kharzeev}\ \emph {et~al.}(2008)\citenamefont
  {Kharzeev}, \citenamefont {McLerran},\ and\ \citenamefont
  {Warringa}}]{Kharzeev:2007jp}%
  \BibitemOpen
  \bibfield  {author} {\bibinfo {author} {\bibfnamefont {Dmitri~E.}\
  \bibnamefont {Kharzeev}}, \bibinfo {author} {\bibfnamefont {Larry~D.}\
  \bibnamefont {McLerran}}, \ and\ \bibinfo {author} {\bibfnamefont
  {Harmen~J.}\ \bibnamefont {Warringa}},\ }\bibfield  {title} {\enquote
  {\bibinfo {title} {{The Effects of topological charge change in heavy ion
  collisions: 'Event by event P and CP violation'}},}\ }\href {\doibase
  10.1016/j.nuclphysa.2008.02.298} {\bibfield  {journal} {\bibinfo  {journal}
  {Nucl.Phys.}\ }\textbf {\bibinfo {volume} {A803}},\ \bibinfo {pages}
  {227--253} (\bibinfo {year} {2008})},\ \Eprint
  {http://arxiv.org/abs/0711.0950} {arXiv:0711.0950 [hep-ph]} \BibitemShut
  {NoStop}%
\bibitem [{\citenamefont {Fukushima}\ \emph {et~al.}(2008)\citenamefont
  {Fukushima}, \citenamefont {Kharzeev},\ and\ \citenamefont
  {Warringa}}]{Fukushima:2008xe}%
  \BibitemOpen
  \bibfield  {author} {\bibinfo {author} {\bibfnamefont {Kenji}\ \bibnamefont
  {Fukushima}}, \bibinfo {author} {\bibfnamefont {Dmitri~E.}\ \bibnamefont
  {Kharzeev}}, \ and\ \bibinfo {author} {\bibfnamefont {Harmen~J.}\
  \bibnamefont {Warringa}},\ }\bibfield  {title} {\enquote {\bibinfo {title}
  {{The Chiral Magnetic Effect}},}\ }\href {\doibase
  10.1103/PhysRevD.78.074033} {\bibfield  {journal} {\bibinfo  {journal} {Phys.
  Rev.}\ }\textbf {\bibinfo {volume} {D78}},\ \bibinfo {pages} {074033}
  (\bibinfo {year} {2008})},\ \Eprint {http://arxiv.org/abs/0808.3382}
  {arXiv:0808.3382 [hep-ph]} \BibitemShut {NoStop}%
\bibitem [{\citenamefont {Kharzeev}\ \emph {et~al.}(2016)\citenamefont
  {Kharzeev}, \citenamefont {Liao}, \citenamefont {Voloshin},\ and\
  \citenamefont {Wang}}]{Kharzeev:2015znc}%
  \BibitemOpen
  \bibfield  {author} {\bibinfo {author} {\bibfnamefont {D.~E.}\ \bibnamefont
  {Kharzeev}}, \bibinfo {author} {\bibfnamefont {J.}~\bibnamefont {Liao}},
  \bibinfo {author} {\bibfnamefont {S.~A.}\ \bibnamefont {Voloshin}}, \ and\
  \bibinfo {author} {\bibfnamefont {G.}~\bibnamefont {Wang}},\ }\bibfield
  {title} {\enquote {\bibinfo {title} {{Chiral magnetic and vortical effects in
  high-energy nuclear collisions---status report}},}\ }\href {\doibase
  10.1016/j.ppnp.2016.01.001} {\bibfield  {journal} {\bibinfo  {journal} {Prog.
  Part. Nucl. Phys.}\ }\textbf {\bibinfo {volume} {88}},\ \bibinfo {pages}
  {1--28} (\bibinfo {year} {2016})},\ \Eprint {http://arxiv.org/abs/1511.04050}
  {arXiv:1511.04050 [hep-ph]} \BibitemShut {NoStop}%
\bibitem [{\citenamefont {Skokov}\ \emph {et~al.}(2017)\citenamefont {Skokov},
  \citenamefont {Sorensen}, \citenamefont {Koch}, \citenamefont {Schlichting},
  \citenamefont {Thomas}, \citenamefont {Voloshin}, \citenamefont {Wang},\ and\
  \citenamefont {Yee}}]{Skokov:2016yrj}%
  \BibitemOpen
  \bibfield  {author} {\bibinfo {author} {\bibfnamefont {Vladimir}\
  \bibnamefont {Skokov}}, \bibinfo {author} {\bibfnamefont {Paul}\ \bibnamefont
  {Sorensen}}, \bibinfo {author} {\bibfnamefont {Volker}\ \bibnamefont {Koch}},
  \bibinfo {author} {\bibfnamefont {Soeren}\ \bibnamefont {Schlichting}},
  \bibinfo {author} {\bibfnamefont {Jim}\ \bibnamefont {Thomas}}, \bibinfo
  {author} {\bibfnamefont {Sergei}\ \bibnamefont {Voloshin}}, \bibinfo {author}
  {\bibfnamefont {Gang}\ \bibnamefont {Wang}}, \ and\ \bibinfo {author}
  {\bibfnamefont {Ho-Ung}\ \bibnamefont {Yee}},\ }\bibfield  {title} {\enquote
  {\bibinfo {title} {{Chiral Magnetic Effect Task Force Report}},}\ }\href
  {\doibase 10.1088/1674-1137/41/7/072001} {\bibfield  {journal} {\bibinfo
  {journal} {Chin. Phys.}\ }\textbf {\bibinfo {volume} {C41}},\ \bibinfo
  {pages} {072001} (\bibinfo {year} {2017})},\ \Eprint
  {http://arxiv.org/abs/1608.00982} {arXiv:1608.00982 [nucl-th]} \BibitemShut
  {NoStop}%
\bibitem [{\citenamefont {Hattori}\ and\ \citenamefont
  {Huang}(2017)}]{Hattori:2016emy}%
  \BibitemOpen
  \bibfield  {author} {\bibinfo {author} {\bibfnamefont {Koichi}\ \bibnamefont
  {Hattori}}\ and\ \bibinfo {author} {\bibfnamefont {Xu-Guang}\ \bibnamefont
  {Huang}},\ }\bibfield  {title} {\enquote {\bibinfo {title} {{Novel quantum
  phenomena induced by strong magnetic fields in heavy-ion collisions}},}\
  }\href {\doibase 10.1007/s41365-016-0178-3} {\bibfield  {journal} {\bibinfo
  {journal} {Nucl. Sci. Tech.}\ }\textbf {\bibinfo {volume} {28}},\ \bibinfo
  {pages} {26} (\bibinfo {year} {2017})},\ \Eprint
  {http://arxiv.org/abs/1609.00747} {arXiv:1609.00747 [nucl-th]} \BibitemShut
  {NoStop}%
\bibitem [{\citenamefont {Burkov}(2015)}]{Burkov:2015hba}%
  \BibitemOpen
  \bibfield  {author} {\bibinfo {author} {\bibfnamefont {A.~A.}\ \bibnamefont
  {Burkov}},\ }\bibfield  {title} {\enquote {\bibinfo {title} {{Chiral anomaly
  and transport in Weyl metals}},}\ }\href {\doibase
  10.1088/0953-8984/27/11/113201} {\bibfield  {journal} {\bibinfo  {journal}
  {J. Phys. Condens. Matter}\ }\textbf {\bibinfo {volume} {27}},\ \bibinfo
  {pages} {113201} (\bibinfo {year} {2015})},\ \Eprint
  {http://arxiv.org/abs/1502.07609} {arXiv:1502.07609 [cond-mat.mes-hall]}
  \BibitemShut {NoStop}%
\bibitem [{\citenamefont {Gorbar}\ \emph {et~al.}(2018)\citenamefont {Gorbar},
  \citenamefont {Miransky}, \citenamefont {Shovkovy},\ and\ \citenamefont
  {Sukhachov}}]{Gorbar:2017lnp}%
  \BibitemOpen
  \bibfield  {author} {\bibinfo {author} {\bibfnamefont {E.~V.}\ \bibnamefont
  {Gorbar}}, \bibinfo {author} {\bibfnamefont {V.~A.}\ \bibnamefont
  {Miransky}}, \bibinfo {author} {\bibfnamefont {I.~A.}\ \bibnamefont
  {Shovkovy}}, \ and\ \bibinfo {author} {\bibfnamefont {P.~O.}\ \bibnamefont
  {Sukhachov}},\ }\bibfield  {title} {\enquote {\bibinfo {title} {{Anomalous
  transport properties of Dirac and Weyl semimetals (Review Article)}},}\
  }\href {\doibase 10.1063/1.5037551} {\bibfield  {journal} {\bibinfo
  {journal} {Low Temp. Phys.}\ }\textbf {\bibinfo {volume} {44}},\ \bibinfo
  {pages} {487--505} (\bibinfo {year} {2018})},\ \Eprint
  {http://arxiv.org/abs/1712.08947} {arXiv:1712.08947 [cond-mat.mes-hall]}
  \BibitemShut {NoStop}%
\bibitem [{\citenamefont {{Armitage}}\ \emph {et~al.}(2018)\citenamefont
  {{Armitage}}, \citenamefont {{Mele}},\ and\ \citenamefont
  {{Vishwanath}}}]{2018RvMP...90a5001A}%
  \BibitemOpen
  \bibfield  {author} {\bibinfo {author} {\bibfnamefont {N.~P.}\ \bibnamefont
  {{Armitage}}}, \bibinfo {author} {\bibfnamefont {E.~J.}\ \bibnamefont
  {{Mele}}}, \ and\ \bibinfo {author} {\bibfnamefont {Ashvin}\ \bibnamefont
  {{Vishwanath}}},\ }\bibfield  {title} {\enquote {\bibinfo {title} {{Weyl and
  Dirac semimetals in three-dimensional solids}},}\ }\href {\doibase
  10.1103/RevModPhys.90.015001} {\bibfield  {journal} {\bibinfo  {journal}
  {Reviews of Modern Physics}\ }\textbf {\bibinfo {volume} {90}},\ \bibinfo
  {eid} {015001} (\bibinfo {year} {2018})},\ \Eprint
  {http://arxiv.org/abs/1705.01111} {arXiv:1705.01111 [cond-mat.str-el]}
  \BibitemShut {NoStop}%
\bibitem [{\citenamefont {Akamatsu}\ and\ \citenamefont
  {Yamamoto}(2013)}]{Akamatsu:2013pjd}%
  \BibitemOpen
  \bibfield  {author} {\bibinfo {author} {\bibfnamefont {Yukinao}\ \bibnamefont
  {Akamatsu}}\ and\ \bibinfo {author} {\bibfnamefont {Naoki}\ \bibnamefont
  {Yamamoto}},\ }\bibfield  {title} {\enquote {\bibinfo {title} {{Chiral Plasma
  Instabilities}},}\ }\href {\doibase 10.1103/PhysRevLett.111.052002}
  {\bibfield  {journal} {\bibinfo  {journal} {Phys. Rev. Lett.}\ }\textbf
  {\bibinfo {volume} {111}},\ \bibinfo {pages} {052002} (\bibinfo {year}
  {2013})},\ \Eprint {http://arxiv.org/abs/1302.2125} {arXiv:1302.2125
  [nucl-th]} \BibitemShut {NoStop}%
\bibitem [{\citenamefont {Yamamoto}(2016)}]{Yamamoto:2015gzz}%
  \BibitemOpen
  \bibfield  {author} {\bibinfo {author} {\bibfnamefont {Naoki}\ \bibnamefont
  {Yamamoto}},\ }\bibfield  {title} {\enquote {\bibinfo {title} {{Chiral
  transport of neutrinos in supernovae: Neutrino-induced fluid helicity and
  helical plasma instability}},}\ }\href {\doibase 10.1103/PhysRevD.93.065017}
  {\bibfield  {journal} {\bibinfo  {journal} {Phys. Rev. D}\ }\textbf {\bibinfo
  {volume} {93}},\ \bibinfo {pages} {065017} (\bibinfo {year} {2016})},\
  \Eprint {http://arxiv.org/abs/1511.00933} {arXiv:1511.00933 [astro-ph.HE]}
  \BibitemShut {NoStop}%
\bibitem [{\citenamefont {Yamamoto}\ and\ \citenamefont
  {Yang}(2020)}]{Yamamoto:2020zrs}%
  \BibitemOpen
  \bibfield  {author} {\bibinfo {author} {\bibfnamefont {Naoki}\ \bibnamefont
  {Yamamoto}}\ and\ \bibinfo {author} {\bibfnamefont {Di-Lun}\ \bibnamefont
  {Yang}},\ }\bibfield  {title} {\enquote {\bibinfo {title} {{Chiral Radiation
  Transport Theory of Neutrinos}},}\ }\href {\doibase 10.3847/1538-4357/ab8468}
  {\bibfield  {journal} {\bibinfo  {journal} {Astrophys. J.}\ }\textbf
  {\bibinfo {volume} {895}},\ \bibinfo {pages} {56} (\bibinfo {year} {2020})},\
  \Eprint {http://arxiv.org/abs/2002.11348} {arXiv:2002.11348 [astro-ph.HE]}
  \BibitemShut {NoStop}%
\bibitem [{\citenamefont {Turner}\ and\ \citenamefont
  {Widrow}(1988)}]{Turner:1987bw}%
  \BibitemOpen
  \bibfield  {author} {\bibinfo {author} {\bibfnamefont {Michael~S.}\
  \bibnamefont {Turner}}\ and\ \bibinfo {author} {\bibfnamefont {Lawrence~M.}\
  \bibnamefont {Widrow}},\ }\bibfield  {title} {\enquote {\bibinfo {title}
  {{Inflation Produced, Large Scale Magnetic Fields}},}\ }\href {\doibase
  10.1103/PhysRevD.37.2743} {\bibfield  {journal} {\bibinfo  {journal} {Phys.
  Rev.}\ }\textbf {\bibinfo {volume} {D37}},\ \bibinfo {pages} {2743} (\bibinfo
  {year} {1988})}\BibitemShut {NoStop}%
\bibitem [{\citenamefont {Carroll}\ \emph {et~al.}(1990)\citenamefont
  {Carroll}, \citenamefont {Field},\ and\ \citenamefont
  {Jackiw}}]{Carroll:1989vb}%
  \BibitemOpen
  \bibfield  {author} {\bibinfo {author} {\bibfnamefont {Sean~M.}\ \bibnamefont
  {Carroll}}, \bibinfo {author} {\bibfnamefont {George~B.}\ \bibnamefont
  {Field}}, \ and\ \bibinfo {author} {\bibfnamefont {Roman}\ \bibnamefont
  {Jackiw}},\ }\bibfield  {title} {\enquote {\bibinfo {title} {{Limits on a
  Lorentz and Parity Violating Modification of Electrodynamics}},}\ }\href
  {\doibase 10.1103/PhysRevD.41.1231} {\bibfield  {journal} {\bibinfo
  {journal} {Phys. Rev.}\ }\textbf {\bibinfo {volume} {D41}},\ \bibinfo {pages}
  {1231} (\bibinfo {year} {1990})}\BibitemShut {NoStop}%
\bibitem [{\citenamefont {Garretson}\ \emph {et~al.}(1992)\citenamefont
  {Garretson}, \citenamefont {Field},\ and\ \citenamefont
  {Carroll}}]{Garretson:1992vt}%
  \BibitemOpen
  \bibfield  {author} {\bibinfo {author} {\bibfnamefont {W.~Daniel}\
  \bibnamefont {Garretson}}, \bibinfo {author} {\bibfnamefont {George~B.}\
  \bibnamefont {Field}}, \ and\ \bibinfo {author} {\bibfnamefont {Sean~M.}\
  \bibnamefont {Carroll}},\ }\bibfield  {title} {\enquote {\bibinfo {title}
  {{Primordial magnetic fields from pseudoGoldstone bosons}},}\ }\href
  {\doibase 10.1103/PhysRevD.46.5346} {\bibfield  {journal} {\bibinfo
  {journal} {Phys. Rev.}\ }\textbf {\bibinfo {volume} {D46}},\ \bibinfo {pages}
  {5346--5351} (\bibinfo {year} {1992})},\ \Eprint
  {http://arxiv.org/abs/hep-ph/9209238} {arXiv:hep-ph/9209238 [hep-ph]}
  \BibitemShut {NoStop}%
\bibitem [{\citenamefont {Joyce}\ and\ \citenamefont
  {Shaposhnikov}(1997)}]{Joyce:1997uy}%
  \BibitemOpen
  \bibfield  {author} {\bibinfo {author} {\bibfnamefont {M.}~\bibnamefont
  {Joyce}}\ and\ \bibinfo {author} {\bibfnamefont {Mikhail~E.}\ \bibnamefont
  {Shaposhnikov}},\ }\bibfield  {title} {\enquote {\bibinfo {title}
  {{Primordial magnetic fields, right-handed electrons, and the Abelian
  anomaly}},}\ }\href {\doibase 10.1103/PhysRevLett.79.1193} {\bibfield
  {journal} {\bibinfo  {journal} {Phys. Rev. Lett.}\ }\textbf {\bibinfo
  {volume} {79}},\ \bibinfo {pages} {1193--1196} (\bibinfo {year} {1997})},\
  \Eprint {http://arxiv.org/abs/astro-ph/9703005} {arXiv:astro-ph/9703005
  [astro-ph]} \BibitemShut {NoStop}%
\bibitem [{\citenamefont {Giovannini}(2004)}]{Giovannini:2003yn}%
  \BibitemOpen
  \bibfield  {author} {\bibinfo {author} {\bibfnamefont {Massimo}\ \bibnamefont
  {Giovannini}},\ }\bibfield  {title} {\enquote {\bibinfo {title} {{The
  Magnetized universe}},}\ }\href {\doibase 10.1142/S0218271804004530}
  {\bibfield  {journal} {\bibinfo  {journal} {Int. J. Mod. Phys. D}\ }\textbf
  {\bibinfo {volume} {13}},\ \bibinfo {pages} {391--502} (\bibinfo {year}
  {2004})},\ \Eprint {http://arxiv.org/abs/astro-ph/0312614}
  {arXiv:astro-ph/0312614} \BibitemShut {NoStop}%
\bibitem [{\citenamefont {Laine}(2005)}]{Laine:2005bt}%
  \BibitemOpen
  \bibfield  {author} {\bibinfo {author} {\bibfnamefont {M.}~\bibnamefont
  {Laine}},\ }\bibfield  {title} {\enquote {\bibinfo {title} {{Real-time
  Chern-Simons term for hypermagnetic fields}},}\ }\href {\doibase
  10.1088/1126-6708/2005/10/056} {\bibfield  {journal} {\bibinfo  {journal}
  {JHEP}\ }\textbf {\bibinfo {volume} {10}},\ \bibinfo {pages} {056} (\bibinfo
  {year} {2005})},\ \Eprint {http://arxiv.org/abs/hep-ph/0508195}
  {arXiv:hep-ph/0508195} \BibitemShut {NoStop}%
\bibitem [{\citenamefont {Abdallah}\ \emph {et~al.}(2022)\citenamefont
  {Abdallah} \emph {et~al.}}]{STAR:2021mii}%
  \BibitemOpen
  \bibfield  {author} {\bibinfo {author} {\bibfnamefont {Mohamed}\ \bibnamefont
  {Abdallah}} \emph {et~al.} (\bibinfo {collaboration} {STAR}),\ }\bibfield
  {title} {\enquote {\bibinfo {title} {{Search for the chiral magnetic effect
  with isobar collisions at $\sqrt {s_{NN}}$=200 GeV by the STAR Collaboration
  at the BNL Relativistic Heavy Ion Collider}},}\ }\href {\doibase
  10.1103/PhysRevC.105.014901} {\bibfield  {journal} {\bibinfo  {journal}
  {Phys. Rev. C}\ }\textbf {\bibinfo {volume} {105}},\ \bibinfo {pages}
  {014901} (\bibinfo {year} {2022})},\ \Eprint
  {http://arxiv.org/abs/2109.00131} {arXiv:2109.00131 [nucl-ex]} \BibitemShut
  {NoStop}%
\bibitem [{\citenamefont {Tanji}\ \emph {et~al.}(2016)\citenamefont {Tanji},
  \citenamefont {Mueller},\ and\ \citenamefont {Berges}}]{Tanji:2016dka}%
  \BibitemOpen
  \bibfield  {author} {\bibinfo {author} {\bibfnamefont {Naoto}\ \bibnamefont
  {Tanji}}, \bibinfo {author} {\bibfnamefont {Niklas}\ \bibnamefont {Mueller}},
  \ and\ \bibinfo {author} {\bibfnamefont {Jurgen}\ \bibnamefont {Berges}},\
  }\bibfield  {title} {\enquote {\bibinfo {title} {{Transient anomalous charge
  production in strong-field QCD}},}\ }\href {\doibase
  10.1103/PhysRevD.93.074507} {\bibfield  {journal} {\bibinfo  {journal} {Phys.
  Rev.}\ }\textbf {\bibinfo {volume} {D93}},\ \bibinfo {pages} {074507}
  (\bibinfo {year} {2016})},\ \Eprint {http://arxiv.org/abs/1603.03331}
  {arXiv:1603.03331 [hep-ph]} \BibitemShut {NoStop}%
\bibitem [{\citenamefont {Mueller}\ \emph {et~al.}(2016)\citenamefont
  {Mueller}, \citenamefont {Schlichting},\ and\ \citenamefont
  {Sharma}}]{Mueller:2016ven}%
  \BibitemOpen
  \bibfield  {author} {\bibinfo {author} {\bibfnamefont {Niklas}\ \bibnamefont
  {Mueller}}, \bibinfo {author} {\bibfnamefont {Soren}\ \bibnamefont
  {Schlichting}}, \ and\ \bibinfo {author} {\bibfnamefont {Sayantan}\
  \bibnamefont {Sharma}},\ }\bibfield  {title} {\enquote {\bibinfo {title}
  {{Chiral magnetic effect and anomalous transport from real-time lattice
  simulations}},}\ }\href {\doibase 10.1103/PhysRevLett.117.142301} {\bibfield
  {journal} {\bibinfo  {journal} {Phys. Rev. Lett.}\ }\textbf {\bibinfo
  {volume} {117}},\ \bibinfo {pages} {142301} (\bibinfo {year} {2016})},\
  \Eprint {http://arxiv.org/abs/1606.00342} {arXiv:1606.00342 [hep-ph]}
  \BibitemShut {NoStop}%
\bibitem [{\citenamefont {Hou}\ and\ \citenamefont {Lin}(2018)}]{Hou:2017szz}%
  \BibitemOpen
  \bibfield  {author} {\bibinfo {author} {\bibfnamefont {De-fu}\ \bibnamefont
  {Hou}}\ and\ \bibinfo {author} {\bibfnamefont {Shu}\ \bibnamefont {Lin}},\
  }\bibfield  {title} {\enquote {\bibinfo {title} {{Fluctuation and Dissipation
  of Axial Charge from Massive Quarks}},}\ }\href {\doibase
  10.1103/PhysRevD.98.054014} {\bibfield  {journal} {\bibinfo  {journal} {Phys.
  Rev.}\ }\textbf {\bibinfo {volume} {D98}},\ \bibinfo {pages} {054014}
  (\bibinfo {year} {2018})},\ \Eprint {http://arxiv.org/abs/1712.08429}
  {arXiv:1712.08429 [hep-ph]} \BibitemShut {NoStop}%
\bibitem [{\citenamefont {Tanji}(2018)}]{Tanji:2018qws}%
  \BibitemOpen
  \bibfield  {author} {\bibinfo {author} {\bibfnamefont {Naoto}\ \bibnamefont
  {Tanji}},\ }\bibfield  {title} {\enquote {\bibinfo {title} {{Nonequilibrium
  axial charge production in expanding glasma flux tubes}},}\ }\href {\doibase
  10.1103/PhysRevD.98.014025} {\bibfield  {journal} {\bibinfo  {journal} {Phys.
  Rev.}\ }\textbf {\bibinfo {volume} {D98}},\ \bibinfo {pages} {014025}
  (\bibinfo {year} {2018})},\ \Eprint {http://arxiv.org/abs/1805.00775}
  {arXiv:1805.00775 [hep-ph]} \BibitemShut {NoStop}%
\bibitem [{\citenamefont {Copinger}\ \emph {et~al.}(2018)\citenamefont
  {Copinger}, \citenamefont {Fukushima},\ and\ \citenamefont
  {Pu}}]{Copinger:2018ftr}%
  \BibitemOpen
  \bibfield  {author} {\bibinfo {author} {\bibfnamefont {Patrick}\ \bibnamefont
  {Copinger}}, \bibinfo {author} {\bibfnamefont {Kenji}\ \bibnamefont
  {Fukushima}}, \ and\ \bibinfo {author} {\bibfnamefont {Shi}\ \bibnamefont
  {Pu}},\ }\bibfield  {title} {\enquote {\bibinfo {title} {{Axial Ward identity
  and the Schwinger mechanism -- Applications to the real-time chiral magnetic
  effect and condensates}},}\ }\href {\doibase 10.1103/PhysRevLett.121.261602}
  {\bibfield  {journal} {\bibinfo  {journal} {Phys. Rev. Lett.}\ }\textbf
  {\bibinfo {volume} {121}},\ \bibinfo {pages} {261602} (\bibinfo {year}
  {2018})},\ \Eprint {http://arxiv.org/abs/1807.04416} {arXiv:1807.04416
  [hep-th]} \BibitemShut {NoStop}%
\bibitem [{\citenamefont {Taya}(2020)}]{Taya:2020bcd}%
  \BibitemOpen
  \bibfield  {author} {\bibinfo {author} {\bibfnamefont {Hidetoshi}\
  \bibnamefont {Taya}},\ }\bibfield  {title} {\enquote {\bibinfo {title}
  {{Dynamically assisted Schwinger mechanism and chirality production in
  parallel electromagnetic field}},}\ }\href {\doibase
  10.1103/PhysRevResearch.2.023257} {\bibfield  {journal} {\bibinfo  {journal}
  {Phys. Rev. Res.}\ }\textbf {\bibinfo {volume} {2}},\ \bibinfo {pages}
  {023257} (\bibinfo {year} {2020})},\ \Eprint
  {http://arxiv.org/abs/2003.08948} {arXiv:2003.08948 [hep-ph]} \BibitemShut
  {NoStop}%
\bibitem [{\citenamefont {M\"uller}\ and\ \citenamefont
  {Yang}(2022)}]{Muller:2021hpe}%
  \BibitemOpen
  \bibfield  {author} {\bibinfo {author} {\bibfnamefont {Berndt}\ \bibnamefont
  {M\"uller}}\ and\ \bibinfo {author} {\bibfnamefont {Di-Lun}\ \bibnamefont
  {Yang}},\ }\bibfield  {title} {\enquote {\bibinfo {title} {{Anomalous spin
  polarization from turbulent color fields}},}\ }\href {\doibase
  10.1103/PhysRevD.105.L011901} {\bibfield  {journal} {\bibinfo  {journal}
  {Phys. Rev. D}\ }\textbf {\bibinfo {volume} {105}},\ \bibinfo {pages}
  {L011901} (\bibinfo {year} {2022})},\ \Eprint
  {http://arxiv.org/abs/2110.15630} {arXiv:2110.15630 [nucl-th]} \BibitemShut
  {NoStop}%
\bibitem [{\citenamefont {Yang}(2021)}]{Yang:2021fea}%
  \BibitemOpen
  \bibfield  {author} {\bibinfo {author} {\bibfnamefont {Di-Lun}\ \bibnamefont
  {Yang}},\ }\bibfield  {title} {\enquote {\bibinfo {title} {{Quantum kinetic
  theory for spin transport of quarks with background chromo-electromagnetic
  fields}},}\ }\href@noop {} {\  (\bibinfo {year} {2021})},\ \Eprint
  {http://arxiv.org/abs/2112.14392} {arXiv:2112.14392 [hep-ph]} \BibitemShut
  {NoStop}%
\bibitem [{\citenamefont {Grabowska}\ \emph {et~al.}(2015)\citenamefont
  {Grabowska}, \citenamefont {Kaplan},\ and\ \citenamefont
  {Reddy}}]{Grabowska:2014efa}%
  \BibitemOpen
  \bibfield  {author} {\bibinfo {author} {\bibfnamefont {Dorota}\ \bibnamefont
  {Grabowska}}, \bibinfo {author} {\bibfnamefont {David~B.}\ \bibnamefont
  {Kaplan}}, \ and\ \bibinfo {author} {\bibfnamefont {Sanjay}\ \bibnamefont
  {Reddy}},\ }\bibfield  {title} {\enquote {\bibinfo {title} {{Role of the
  electron mass in damping chiral plasma instability in Supernovae and neutron
  stars}},}\ }\href {\doibase 10.1103/PhysRevD.91.085035} {\bibfield  {journal}
  {\bibinfo  {journal} {Phys. Rev.}\ }\textbf {\bibinfo {volume} {D91}},\
  \bibinfo {pages} {085035} (\bibinfo {year} {2015})},\ \Eprint
  {http://arxiv.org/abs/1409.3602} {arXiv:1409.3602 [hep-ph]} \BibitemShut
  {NoStop}%
\bibitem [{\citenamefont {Gorbar}\ and\ \citenamefont
  {Shovkovy}(2021)}]{Gorbar:2021tnw}%
  \BibitemOpen
  \bibfield  {author} {\bibinfo {author} {\bibfnamefont {E.~V.}\ \bibnamefont
  {Gorbar}}\ and\ \bibinfo {author} {\bibfnamefont {I.~A.}\ \bibnamefont
  {Shovkovy}},\ }\bibfield  {title} {\enquote {\bibinfo {title} {{Chiral
  anomalous processes in magnetospheres of pulsars and black holes}},}\
  }\href@noop {} {\  (\bibinfo {year} {2021})},\ \Eprint
  {http://arxiv.org/abs/2110.11380} {arXiv:2110.11380 [astro-ph.HE]}
  \BibitemShut {NoStop}%
\bibitem [{\citenamefont {Domcke}\ \emph {et~al.}(2020)\citenamefont {Domcke},
  \citenamefont {Ema},\ and\ \citenamefont {Mukaida}}]{Domcke:2019qmm}%
  \BibitemOpen
  \bibfield  {author} {\bibinfo {author} {\bibfnamefont {Valerie}\ \bibnamefont
  {Domcke}}, \bibinfo {author} {\bibfnamefont {Yohei}\ \bibnamefont {Ema}}, \
  and\ \bibinfo {author} {\bibfnamefont {Kyohei}\ \bibnamefont {Mukaida}},\
  }\bibfield  {title} {\enquote {\bibinfo {title} {{Chiral Anomaly, Schwinger
  Effect, Euler-Heisenberg Lagrangian, and application to axion inflation}},}\
  }\href {\doibase 10.1007/JHEP02(2020)055} {\bibfield  {journal} {\bibinfo
  {journal} {JHEP}\ }\textbf {\bibinfo {volume} {02}},\ \bibinfo {pages} {055}
  (\bibinfo {year} {2020})},\ \Eprint {http://arxiv.org/abs/1910.01205}
  {arXiv:1910.01205 [hep-ph]} \BibitemShut {NoStop}%
\bibitem [{\citenamefont {Boyarsky}\ \emph {et~al.}(2021)\citenamefont
  {Boyarsky}, \citenamefont {Cheianov}, \citenamefont {Ruchayskiy},\ and\
  \citenamefont {Sobol}}]{Boyarsky:2020cyk}%
  \BibitemOpen
  \bibfield  {author} {\bibinfo {author} {\bibfnamefont {A.}~\bibnamefont
  {Boyarsky}}, \bibinfo {author} {\bibfnamefont {V.}~\bibnamefont {Cheianov}},
  \bibinfo {author} {\bibfnamefont {O.}~\bibnamefont {Ruchayskiy}}, \ and\
  \bibinfo {author} {\bibfnamefont {O.}~\bibnamefont {Sobol}},\ }\bibfield
  {title} {\enquote {\bibinfo {title} {{Evolution of the Primordial Axial
  Charge across Cosmic Times}},}\ }\href {\doibase
  10.1103/PhysRevLett.126.021801} {\bibfield  {journal} {\bibinfo  {journal}
  {Phys. Rev. Lett.}\ }\textbf {\bibinfo {volume} {126}},\ \bibinfo {pages}
  {021801} (\bibinfo {year} {2021})},\ \Eprint
  {http://arxiv.org/abs/2007.13691} {arXiv:2007.13691 [hep-ph]} \BibitemShut
  {NoStop}%
\bibitem [{\citenamefont {Adler}(1969)}]{Adler:1969gk}%
  \BibitemOpen
  \bibfield  {author} {\bibinfo {author} {\bibfnamefont {Stephen~L.}\
  \bibnamefont {Adler}},\ }\bibfield  {title} {\enquote {\bibinfo {title}
  {{Axial vector vertex in spinor electrodynamics}},}\ }\href {\doibase
  10.1103/PhysRev.177.2426} {\bibfield  {journal} {\bibinfo  {journal} {Phys.
  Rev.}\ }\textbf {\bibinfo {volume} {177}},\ \bibinfo {pages} {2426--2438}
  (\bibinfo {year} {1969})}\BibitemShut {NoStop}%
\bibitem [{\citenamefont {Bell}\ and\ \citenamefont
  {Jackiw}(1969)}]{Bell:1969ts}%
  \BibitemOpen
  \bibfield  {author} {\bibinfo {author} {\bibfnamefont {J.~S.}\ \bibnamefont
  {Bell}}\ and\ \bibinfo {author} {\bibfnamefont {R.}~\bibnamefont {Jackiw}},\
  }\bibfield  {title} {\enquote {\bibinfo {title} {{A PCAC puzzle: $\pi^0 \to
  \gamma \gamma$ in the sigma model}},}\ }\href {\doibase 10.1007/BF02823296}
  {\bibfield  {journal} {\bibinfo  {journal} {Nuovo Cim.}\ }\textbf {\bibinfo
  {volume} {A60}},\ \bibinfo {pages} {47--61} (\bibinfo {year}
  {1969})}\BibitemShut {NoStop}%
\bibitem [{\citenamefont {Adler}\ and\ \citenamefont
  {Bardeen}(1969)}]{Adler:1969er}%
  \BibitemOpen
  \bibfield  {author} {\bibinfo {author} {\bibfnamefont {Stephen~L.}\
  \bibnamefont {Adler}}\ and\ \bibinfo {author} {\bibfnamefont {William~A.}\
  \bibnamefont {Bardeen}},\ }\bibfield  {title} {\enquote {\bibinfo {title}
  {{Absence of higher order corrections in the anomalous axial vector
  divergence equation}},}\ }\href {\doibase 10.1103/PhysRev.182.1517}
  {\bibfield  {journal} {\bibinfo  {journal} {Phys. Rev.}\ }\textbf {\bibinfo
  {volume} {182}},\ \bibinfo {pages} {1517--1536} (\bibinfo {year}
  {1969})}\BibitemShut {NoStop}%
\bibitem [{\citenamefont {Itoyama}\ and\ \citenamefont
  {Mueller}(1983)}]{Itoyama:1982up}%
  \BibitemOpen
  \bibfield  {author} {\bibinfo {author} {\bibfnamefont {Hiroshi}\ \bibnamefont
  {Itoyama}}\ and\ \bibinfo {author} {\bibfnamefont {Alfred~H.}\ \bibnamefont
  {Mueller}},\ }\bibfield  {title} {\enquote {\bibinfo {title} {{The Axial
  Anomaly at Finite Temperature}},}\ }\href {\doibase
  10.1016/0550-3213(83)90370-X} {\bibfield  {journal} {\bibinfo  {journal}
  {Nucl. Phys. B}\ }\textbf {\bibinfo {volume} {218}},\ \bibinfo {pages}
  {349--365} (\bibinfo {year} {1983})}\BibitemShut {NoStop}%
\bibitem [{\citenamefont {Nielsen}\ and\ \citenamefont
  {Ninomiya}(1983)}]{Nielsen:1983rb}%
  \BibitemOpen
  \bibfield  {author} {\bibinfo {author} {\bibfnamefont {Holger~Bech}\
  \bibnamefont {Nielsen}}\ and\ \bibinfo {author} {\bibfnamefont {Masao}\
  \bibnamefont {Ninomiya}},\ }\bibfield  {title} {\enquote {\bibinfo {title}
  {{The Adler-Bell-Jackiw anomaly and Weyl fermions in a crystal}},}\ }\href
  {\doibase 10.1016/0370-2693(83)91529-0} {\bibfield  {journal} {\bibinfo
  {journal} {Phys. Lett.}\ }\textbf {\bibinfo {volume} {130B}},\ \bibinfo
  {pages} {389--396} (\bibinfo {year} {1983})}\BibitemShut {NoStop}%
\bibitem [{\citenamefont {Ambjorn}\ \emph {et~al.}(1983)\citenamefont
  {Ambjorn}, \citenamefont {Greensite},\ and\ \citenamefont
  {Peterson}}]{Ambjorn:1983hp}%
  \BibitemOpen
  \bibfield  {author} {\bibinfo {author} {\bibfnamefont {Jan}\ \bibnamefont
  {Ambjorn}}, \bibinfo {author} {\bibfnamefont {J.}~\bibnamefont {Greensite}},
  \ and\ \bibinfo {author} {\bibfnamefont {C.}~\bibnamefont {Peterson}},\
  }\bibfield  {title} {\enquote {\bibinfo {title} {{The Axial Anomaly and the
  Lattice Dirac Sea}},}\ }\href {\doibase 10.1016/0550-3213(83)90585-0}
  {\bibfield  {journal} {\bibinfo  {journal} {Nucl. Phys.}\ }\textbf {\bibinfo
  {volume} {B221}},\ \bibinfo {pages} {381--408} (\bibinfo {year}
  {1983})}\BibitemShut {NoStop}%
\bibitem [{\citenamefont {Creutz}(2001)}]{Creutz:2000bs}%
  \BibitemOpen
  \bibfield  {author} {\bibinfo {author} {\bibfnamefont {Michael}\ \bibnamefont
  {Creutz}},\ }\bibfield  {title} {\enquote {\bibinfo {title} {{Aspects of
  Chiral Symmetry and the Lattice}},}\ }\href {\doibase
  10.1103/RevModPhys.73.119} {\bibfield  {journal} {\bibinfo  {journal} {Rev.
  Mod. Phys.}\ }\textbf {\bibinfo {volume} {73}},\ \bibinfo {pages} {119--150}
  (\bibinfo {year} {2001})},\ \Eprint {http://arxiv.org/abs/hep-lat/0007032}
  {arXiv:hep-lat/0007032 [hep-lat]} \BibitemShut {NoStop}%
\bibitem [{\citenamefont {Peskin}\ and\ \citenamefont
  {Schroeder}(1995)}]{Peskin:1995ev}%
  \BibitemOpen
  \bibfield  {author} {\bibinfo {author} {\bibfnamefont {Michael~E.}\
  \bibnamefont {Peskin}}\ and\ \bibinfo {author} {\bibfnamefont {Daniel~V.}\
  \bibnamefont {Schroeder}},\ }\href
  {http://www.slac.stanford.edu/spires/find/books/www?cl=QC174.45%3AP4} {\emph
  {\bibinfo {title} {{An Introduction to quantum field theory}}}}\ (\bibinfo
  {year} {1995})\BibitemShut {NoStop}%
\bibitem [{\citenamefont {Hattori}\ and\ \citenamefont
  {Itakura}(2013)}]{Hattori:2012je}%
  \BibitemOpen
  \bibfield  {author} {\bibinfo {author} {\bibfnamefont {Koichi}\ \bibnamefont
  {Hattori}}\ and\ \bibinfo {author} {\bibfnamefont {Kazunori}\ \bibnamefont
  {Itakura}},\ }\bibfield  {title} {\enquote {\bibinfo {title} {{Vacuum
  birefringence in strong magnetic fields: (I) Photon polarization tensor with
  all the Landau levels}},}\ }\href {\doibase 10.1016/j.aop.2012.11.010}
  {\bibfield  {journal} {\bibinfo  {journal} {Annals Phys.}\ }\textbf {\bibinfo
  {volume} {330}},\ \bibinfo {pages} {23--54} (\bibinfo {year} {2013})},\
  \Eprint {http://arxiv.org/abs/1209.2663} {arXiv:1209.2663 [hep-ph]}
  \BibitemShut {NoStop}%
\bibitem [{\citenamefont {Hattori}\ and\ \citenamefont
  {Satow}(2018)}]{Hattori:2017xoo}%
  \BibitemOpen
  \bibfield  {author} {\bibinfo {author} {\bibfnamefont {Koichi}\ \bibnamefont
  {Hattori}}\ and\ \bibinfo {author} {\bibfnamefont {Daisuke}\ \bibnamefont
  {Satow}},\ }\bibfield  {title} {\enquote {\bibinfo {title} {{Gluon Spectrum
  in Quark-Gluon Plasma under Strong Magnetic Fields}},}\ }\href {\doibase
  10.1103/PhysRevD.97.014023} {\bibfield  {journal} {\bibinfo  {journal} {Phys.
  Rev.}\ }\textbf {\bibinfo {volume} {D97}},\ \bibinfo {pages} {014023}
  (\bibinfo {year} {2018})},\ \Eprint {http://arxiv.org/abs/1704.03191}
  {arXiv:1704.03191 [hep-ph]} \BibitemShut {NoStop}%
\bibitem [{\citenamefont {Schwinger}(1962{\natexlab{a}})}]{Schwinger:1962tn}%
  \BibitemOpen
  \bibfield  {author} {\bibinfo {author} {\bibfnamefont {Julian~S.}\
  \bibnamefont {Schwinger}},\ }\bibfield  {title} {\enquote {\bibinfo {title}
  {{Gauge Invariance and Mass}},}\ }\href {\doibase 10.1103/PhysRev.125.397}
  {\bibfield  {journal} {\bibinfo  {journal} {Phys. Rev.}\ }\textbf {\bibinfo
  {volume} {125}},\ \bibinfo {pages} {397--398} (\bibinfo {year}
  {1962}{\natexlab{a}})}\BibitemShut {NoStop}%
\bibitem [{\citenamefont {Schwinger}(1962{\natexlab{b}})}]{Schwinger:1962tp}%
  \BibitemOpen
  \bibfield  {author} {\bibinfo {author} {\bibfnamefont {Julian~S.}\
  \bibnamefont {Schwinger}},\ }\bibfield  {title} {\enquote {\bibinfo {title}
  {{Gauge Invariance and Mass. 2.}}}\ }\href {\doibase
  10.1103/PhysRev.128.2425} {\bibfield  {journal} {\bibinfo  {journal} {Phys.
  Rev.}\ }\textbf {\bibinfo {volume} {128}},\ \bibinfo {pages} {2425--2429}
  (\bibinfo {year} {1962}{\natexlab{b}})}\BibitemShut {NoStop}%
\bibitem [{\citenamefont {Fukushima}\ \emph {et~al.}(2016)\citenamefont
  {Fukushima}, \citenamefont {Hattori}, \citenamefont {Yee},\ and\
  \citenamefont {Yin}}]{Fukushima:2015wck}%
  \BibitemOpen
  \bibfield  {author} {\bibinfo {author} {\bibfnamefont {Kenji}\ \bibnamefont
  {Fukushima}}, \bibinfo {author} {\bibfnamefont {Koichi}\ \bibnamefont
  {Hattori}}, \bibinfo {author} {\bibfnamefont {Ho-Ung}\ \bibnamefont {Yee}}, \
  and\ \bibinfo {author} {\bibfnamefont {Yi}~\bibnamefont {Yin}},\ }\bibfield
  {title} {\enquote {\bibinfo {title} {{Heavy Quark Diffusion in Strong
  Magnetic Fields at Weak Coupling and Implications for Elliptic Flow}},}\
  }\href {\doibase 10.1103/PhysRevD.93.074028} {\bibfield  {journal} {\bibinfo
  {journal} {Phys. Rev.}\ }\textbf {\bibinfo {volume} {D93}},\ \bibinfo {pages}
  {074028} (\bibinfo {year} {2016})},\ \Eprint
  {http://arxiv.org/abs/1512.03689} {arXiv:1512.03689 [hep-ph]} \BibitemShut
  {NoStop}%
\bibitem [{\citenamefont {Smilga}(1992)}]{Smilga:1991xa}%
  \BibitemOpen
  \bibfield  {author} {\bibinfo {author} {\bibfnamefont {Andrei~V.}\
  \bibnamefont {Smilga}},\ }\bibfield  {title} {\enquote {\bibinfo {title}
  {{Anomaly mechanism at finite temperature}},}\ }\href {\doibase
  10.1103/PhysRevD.45.1378} {\bibfield  {journal} {\bibinfo  {journal} {Phys.
  Rev.}\ }\textbf {\bibinfo {volume} {D45}},\ \bibinfo {pages} {1378--1394}
  (\bibinfo {year} {1992})}\BibitemShut {NoStop}%
\bibitem [{\citenamefont {Donoghue}\ \emph {et~al.}(1992)\citenamefont
  {Donoghue}, \citenamefont {Golowich},\ and\ \citenamefont
  {Holstein}}]{Donoghue:1992dd}%
  \BibitemOpen
  \bibfield  {author} {\bibinfo {author} {\bibfnamefont {J.~F.}\ \bibnamefont
  {Donoghue}}, \bibinfo {author} {\bibfnamefont {E.}~\bibnamefont {Golowich}},
  \ and\ \bibinfo {author} {\bibfnamefont {Barry~R.}\ \bibnamefont
  {Holstein}},\ }\bibfield  {title} {\enquote {\bibinfo {title} {{Dynamics of
  the standard model}},}\ }\href {\doibase 10.1017/CBO9780511524370} {\bibfield
   {journal} {\bibinfo  {journal} {Camb. Monogr. Part. Phys. Nucl. Phys.
  Cosmol.}\ }\textbf {\bibinfo {volume} {2}},\ \bibinfo {pages} {1--540}
  (\bibinfo {year} {1992})},\ \bibinfo {note} {[Camb. Monogr. Part. Phys. Nucl.
  Phys. Cosmol.35(2014)]}\BibitemShut {NoStop}%
\bibitem [{\citenamefont {Zyla}\ \emph {et~al.}(2020)\citenamefont {Zyla} \emph
  {et~al.}}]{Zyla:2020zbs}%
  \BibitemOpen
  \bibfield  {author} {\bibinfo {author} {\bibfnamefont {P.A.}\ \bibnamefont
  {Zyla}} \emph {et~al.} (\bibinfo {collaboration} {Particle Data Group}),\
  }\bibfield  {title} {\enquote {\bibinfo {title} {{Review of Particle
  Physics}},}\ }\href {\doibase 10.1093/ptep/ptaa104} {\bibfield  {journal}
  {\bibinfo  {journal} {PTEP}\ }\textbf {\bibinfo {volume} {2020}},\ \bibinfo
  {pages} {083C01} (\bibinfo {year} {2020})}\BibitemShut {NoStop}%
\bibitem [{\citenamefont {Gamow}(1988)}]{gamow1988one}%
  \BibitemOpen
  \bibfield  {author} {\bibinfo {author} {\bibfnamefont {George}\ \bibnamefont
  {Gamow}},\ }\href@noop {} {\emph {\bibinfo {title} {One, two,
  three--infinity: facts and speculations of science}}}\ (\bibinfo  {publisher}
  {Courier Corporation},\ \bibinfo {year} {1988})\BibitemShut {NoStop}%
\bibitem [{\citenamefont {Akamatsu}\ and\ \citenamefont
  {Yamamoto}(2014)}]{Akamatsu:2014yza}%
  \BibitemOpen
  \bibfield  {author} {\bibinfo {author} {\bibfnamefont {Yukinao}\ \bibnamefont
  {Akamatsu}}\ and\ \bibinfo {author} {\bibfnamefont {Naoki}\ \bibnamefont
  {Yamamoto}},\ }\bibfield  {title} {\enquote {\bibinfo {title} {{Chiral
  Langevin theory for non-Abelian plasmas}},}\ }\href {\doibase
  10.1103/PhysRevD.90.125031} {\bibfield  {journal} {\bibinfo  {journal} {Phys.
  Rev.}\ }\textbf {\bibinfo {volume} {D90}},\ \bibinfo {pages} {125031}
  (\bibinfo {year} {2014})},\ \Eprint {http://arxiv.org/abs/1402.4174}
  {arXiv:1402.4174 [hep-th]} \BibitemShut {NoStop}%
\bibitem [{\citenamefont {Zhou}\ \emph {et~al.}(2015)\citenamefont {Zhou},
  \citenamefont {Chang},\ and\ \citenamefont {Xiao}}]{PhysRevB.91.035114}%
  \BibitemOpen
  \bibfield  {author} {\bibinfo {author} {\bibfnamefont {Jianhui}\ \bibnamefont
  {Zhou}}, \bibinfo {author} {\bibfnamefont {Hao-Ran}\ \bibnamefont {Chang}}, \
  and\ \bibinfo {author} {\bibfnamefont {Di}~\bibnamefont {Xiao}},\ }\bibfield
  {title} {\enquote {\bibinfo {title} {Plasmon mode as a detection of the
  chiral anomaly in weyl semimetals},}\ }\href {\doibase
  10.1103/PhysRevB.91.035114} {\bibfield  {journal} {\bibinfo  {journal} {Phys.
  Rev. B}\ }\textbf {\bibinfo {volume} {91}},\ \bibinfo {pages} {035114}
  (\bibinfo {year} {2015})}\BibitemShut {NoStop}%
\bibitem [{\citenamefont {Behrends}\ \emph {et~al.}(2016)\citenamefont
  {Behrends}, \citenamefont {Grushin}, \citenamefont {Ojanen},\ and\
  \citenamefont {Bardarson}}]{PhysRevB.93.075114}%
  \BibitemOpen
  \bibfield  {author} {\bibinfo {author} {\bibfnamefont {Jan}\ \bibnamefont
  {Behrends}}, \bibinfo {author} {\bibfnamefont {Adolfo~G.}\ \bibnamefont
  {Grushin}}, \bibinfo {author} {\bibfnamefont {Teemu}\ \bibnamefont {Ojanen}},
  \ and\ \bibinfo {author} {\bibfnamefont {Jens~H.}\ \bibnamefont
  {Bardarson}},\ }\bibfield  {title} {\enquote {\bibinfo {title} {Visualizing
  the chiral anomaly in dirac and weyl semimetals with photoemission
  spectroscopy},}\ }\href {\doibase 10.1103/PhysRevB.93.075114} {\bibfield
  {journal} {\bibinfo  {journal} {Phys. Rev. B}\ }\textbf {\bibinfo {volume}
  {93}},\ \bibinfo {pages} {075114} (\bibinfo {year} {2016})}\BibitemShut
  {NoStop}%
\bibitem [{\citenamefont {Tabert}\ \emph {et~al.}(2016)\citenamefont {Tabert},
  \citenamefont {Carbotte},\ and\ \citenamefont {Nicol}}]{PhysRevB.93.085426}%
  \BibitemOpen
  \bibfield  {author} {\bibinfo {author} {\bibfnamefont {C.~J.}\ \bibnamefont
  {Tabert}}, \bibinfo {author} {\bibfnamefont {J.~P.}\ \bibnamefont
  {Carbotte}}, \ and\ \bibinfo {author} {\bibfnamefont {E.~J.}\ \bibnamefont
  {Nicol}},\ }\bibfield  {title} {\enquote {\bibinfo {title} {Optical and
  transport properties in three-dimensional dirac and weyl semimetals},}\
  }\href {\doibase 10.1103/PhysRevB.93.085426} {\bibfield  {journal} {\bibinfo
  {journal} {Phys. Rev. B}\ }\textbf {\bibinfo {volume} {93}},\ \bibinfo
  {pages} {085426} (\bibinfo {year} {2016})}\BibitemShut {NoStop}%
\bibitem [{\citenamefont {Yang}\ \emph {et~al.}(2018)\citenamefont {Yang},
  \citenamefont {Kim},\ and\ \citenamefont {Kim}}]{PhysRevB.98.075203}%
  \BibitemOpen
  \bibfield  {author} {\bibinfo {author} {\bibfnamefont {Jinho}\ \bibnamefont
  {Yang}}, \bibinfo {author} {\bibfnamefont {Jeehoon}\ \bibnamefont {Kim}}, \
  and\ \bibinfo {author} {\bibfnamefont {Ki-Seok}\ \bibnamefont {Kim}},\
  }\bibfield  {title} {\enquote {\bibinfo {title} {Transmission and reflection
  coefficients and faraday/kerr rotations as a function of applied magnetic
  fields in spin-orbit coupled dirac metals},}\ }\href {\doibase
  10.1103/PhysRevB.98.075203} {\bibfield  {journal} {\bibinfo  {journal} {Phys.
  Rev. B}\ }\textbf {\bibinfo {volume} {98}},\ \bibinfo {pages} {075203}
  (\bibinfo {year} {2018})}\BibitemShut {NoStop}%
\bibitem [{\citenamefont {Sonowal}\ \emph {et~al.}(2019)\citenamefont
  {Sonowal}, \citenamefont {Singh},\ and\ \citenamefont
  {Agarwal}}]{PhysRevB.100.085436}%
  \BibitemOpen
  \bibfield  {author} {\bibinfo {author} {\bibfnamefont {Kabyashree}\
  \bibnamefont {Sonowal}}, \bibinfo {author} {\bibfnamefont {Ashutosh}\
  \bibnamefont {Singh}}, \ and\ \bibinfo {author} {\bibfnamefont {Amit}\
  \bibnamefont {Agarwal}},\ }\bibfield  {title} {\enquote {\bibinfo {title}
  {Giant optical activity and kerr effect in type-i and type-ii weyl
  semimetals},}\ }\href {\doibase 10.1103/PhysRevB.100.085436} {\bibfield
  {journal} {\bibinfo  {journal} {Phys. Rev. B}\ }\textbf {\bibinfo {volume}
  {100}},\ \bibinfo {pages} {085436} (\bibinfo {year} {2019})}\BibitemShut
  {NoStop}%
\bibitem [{\citenamefont {Song}\ \emph {et~al.}(2016)\citenamefont {Song},
  \citenamefont {Zhao}, \citenamefont {Fang},\ and\ \citenamefont
  {Dai}}]{song2016detecting}%
  \BibitemOpen
  \bibfield  {author} {\bibinfo {author} {\bibfnamefont {Zhida}\ \bibnamefont
  {Song}}, \bibinfo {author} {\bibfnamefont {Jimin}\ \bibnamefont {Zhao}},
  \bibinfo {author} {\bibfnamefont {Zhong}\ \bibnamefont {Fang}}, \ and\
  \bibinfo {author} {\bibfnamefont {Xi}~\bibnamefont {Dai}},\ }\bibfield
  {title} {\enquote {\bibinfo {title} {Detecting the chiral magnetic effect by
  lattice dynamics in weyl semimetals},}\ }\href@noop {} {\bibfield  {journal}
  {\bibinfo  {journal} {Physical Review B}\ }\textbf {\bibinfo {volume} {94}},\
  \bibinfo {pages} {214306} (\bibinfo {year} {2016})}\BibitemShut {NoStop}%
\bibitem [{\citenamefont {Rinkel}\ \emph {et~al.}(2017)\citenamefont {Rinkel},
  \citenamefont {Lopes},\ and\ \citenamefont {Garate}}]{rinkel2017signatures}%
  \BibitemOpen
  \bibfield  {author} {\bibinfo {author} {\bibfnamefont {Pierre}\ \bibnamefont
  {Rinkel}}, \bibinfo {author} {\bibfnamefont {Pedro~LS}\ \bibnamefont
  {Lopes}}, \ and\ \bibinfo {author} {\bibfnamefont {Ion}\ \bibnamefont
  {Garate}},\ }\bibfield  {title} {\enquote {\bibinfo {title} {Signatures of
  the chiral anomaly in phonon dynamics},}\ }\href@noop {} {\bibfield
  {journal} {\bibinfo  {journal} {Physical review letters}\ }\textbf {\bibinfo
  {volume} {119}},\ \bibinfo {pages} {107401} (\bibinfo {year}
  {2017})}\BibitemShut {NoStop}%
\bibitem [{\citenamefont {Hui}\ \emph {et~al.}(2019)\citenamefont {Hui},
  \citenamefont {Zhang},\ and\ \citenamefont {Kim}}]{hui2019optical}%
  \BibitemOpen
  \bibfield  {author} {\bibinfo {author} {\bibfnamefont {Aaron}\ \bibnamefont
  {Hui}}, \bibinfo {author} {\bibfnamefont {Yi}~\bibnamefont {Zhang}}, \ and\
  \bibinfo {author} {\bibfnamefont {Eun-Ah}\ \bibnamefont {Kim}},\ }\bibfield
  {title} {\enquote {\bibinfo {title} {Optical signatures of the chiral anomaly
  in mirror-symmetric weyl semimetals},}\ }\href@noop {} {\bibfield  {journal}
  {\bibinfo  {journal} {Physical Review B}\ }\textbf {\bibinfo {volume}
  {100}},\ \bibinfo {pages} {085144} (\bibinfo {year} {2019})}\BibitemShut
  {NoStop}%
\bibitem [{\citenamefont {Yuan}\ \emph {et~al.}(2020)\citenamefont {Yuan},
  \citenamefont {Zhang}, \citenamefont {Zhang}, \citenamefont {Yan},
  \citenamefont {Lyu}, \citenamefont {Zhang}, \citenamefont {Li}, \citenamefont
  {Song}, \citenamefont {Zhao}, \citenamefont {Leng} \emph
  {et~al.}}]{yuan2020discovery}%
  \BibitemOpen
  \bibfield  {author} {\bibinfo {author} {\bibfnamefont {Xiang}\ \bibnamefont
  {Yuan}}, \bibinfo {author} {\bibfnamefont {Cheng}\ \bibnamefont {Zhang}},
  \bibinfo {author} {\bibfnamefont {Yi}~\bibnamefont {Zhang}}, \bibinfo
  {author} {\bibfnamefont {Zhongbo}\ \bibnamefont {Yan}}, \bibinfo {author}
  {\bibfnamefont {Tairu}\ \bibnamefont {Lyu}}, \bibinfo {author} {\bibfnamefont
  {Mengyao}\ \bibnamefont {Zhang}}, \bibinfo {author} {\bibfnamefont {Zhilin}\
  \bibnamefont {Li}}, \bibinfo {author} {\bibfnamefont {Chaoyu}\ \bibnamefont
  {Song}}, \bibinfo {author} {\bibfnamefont {Minhao}\ \bibnamefont {Zhao}},
  \bibinfo {author} {\bibfnamefont {Pengliang}\ \bibnamefont {Leng}},  \emph
  {et~al.},\ }\bibfield  {title} {\enquote {\bibinfo {title} {The discovery of
  dynamic chiral anomaly in a weyl semimetal nbas},}\ }\href@noop {} {\bibfield
   {journal} {\bibinfo  {journal} {Nature communications}\ }\textbf {\bibinfo
  {volume} {11}},\ \bibinfo {pages} {1--7} (\bibinfo {year}
  {2020})}\BibitemShut {NoStop}%
\bibitem [{\citenamefont {Almutairi}\ \emph {et~al.}(2020)\citenamefont
  {Almutairi}, \citenamefont {Chen}, \citenamefont {Tokman},\ and\
  \citenamefont {Belyanin}}]{PhysRevB.101.235156}%
  \BibitemOpen
  \bibfield  {author} {\bibinfo {author} {\bibfnamefont {Sultan}\ \bibnamefont
  {Almutairi}}, \bibinfo {author} {\bibfnamefont {Qianfan}\ \bibnamefont
  {Chen}}, \bibinfo {author} {\bibfnamefont {Mikhail}\ \bibnamefont {Tokman}},
  \ and\ \bibinfo {author} {\bibfnamefont {Alexey}\ \bibnamefont {Belyanin}},\
  }\bibfield  {title} {\enquote {\bibinfo {title} {Four-wave mixing in {W}eyl
  semimetals},}\ }\href {\doibase 10.1103/PhysRevB.101.235156} {\bibfield
  {journal} {\bibinfo  {journal} {Phys. Rev. B}\ }\textbf {\bibinfo {volume}
  {101}},\ \bibinfo {pages} {235156} (\bibinfo {year} {2020})}\BibitemShut
  {NoStop}%
\bibitem [{\citenamefont {Gao}\ and\ \citenamefont
  {Zhang}(2021)}]{PhysRevB.103.L041301}%
  \BibitemOpen
  \bibfield  {author} {\bibinfo {author} {\bibfnamefont {Yang}\ \bibnamefont
  {Gao}}\ and\ \bibinfo {author} {\bibfnamefont {Furu}\ \bibnamefont {Zhang}},\
  }\bibfield  {title} {\enquote {\bibinfo {title} {Current-induced second
  harmonic generation of dirac or weyl semimetals in a strong magnetic
  field},}\ }\href {\doibase 10.1103/PhysRevB.103.L041301} {\bibfield
  {journal} {\bibinfo  {journal} {Phys. Rev. B}\ }\textbf {\bibinfo {volume}
  {103}},\ \bibinfo {pages} {L041301} (\bibinfo {year} {2021})}\BibitemShut
  {NoStop}%
\bibitem [{\citenamefont {Singh}\ and\ \citenamefont
  {Carbotte}(2021)}]{PhysRevB.103.075114}%
  \BibitemOpen
  \bibfield  {author} {\bibinfo {author} {\bibfnamefont {Ashutosh}\
  \bibnamefont {Singh}}\ and\ \bibinfo {author} {\bibfnamefont {J.~P.}\
  \bibnamefont {Carbotte}},\ }\bibfield  {title} {\enquote {\bibinfo {title}
  {Temperature effects in tilted weyl semimetals: Dichroism and dynamic hall
  angle},}\ }\href {\doibase 10.1103/PhysRevB.103.075114} {\bibfield  {journal}
  {\bibinfo  {journal} {Phys. Rev. B}\ }\textbf {\bibinfo {volume} {103}},\
  \bibinfo {pages} {075114} (\bibinfo {year} {2021})}\BibitemShut {NoStop}%
\bibitem [{\citenamefont {Long}\ \emph {et~al.}(2018)\citenamefont {Long},
  \citenamefont {Wang}, \citenamefont {Erukhimova}, \citenamefont {Tokman},\
  and\ \citenamefont {Belyanin}}]{PhysRevLett.120.037403}%
  \BibitemOpen
  \bibfield  {author} {\bibinfo {author} {\bibfnamefont {Zhongqu}\ \bibnamefont
  {Long}}, \bibinfo {author} {\bibfnamefont {Yongrui}\ \bibnamefont {Wang}},
  \bibinfo {author} {\bibfnamefont {Maria}\ \bibnamefont {Erukhimova}},
  \bibinfo {author} {\bibfnamefont {Mikhail}\ \bibnamefont {Tokman}}, \ and\
  \bibinfo {author} {\bibfnamefont {Alexey}\ \bibnamefont {Belyanin}},\
  }\bibfield  {title} {\enquote {\bibinfo {title} {Magnetopolaritons in weyl
  semimetals in a strong magnetic field},}\ }\href {\doibase
  10.1103/PhysRevLett.120.037403} {\bibfield  {journal} {\bibinfo  {journal}
  {Phys. Rev. Lett.}\ }\textbf {\bibinfo {volume} {120}},\ \bibinfo {pages}
  {037403} (\bibinfo {year} {2018})}\BibitemShut {NoStop}%
\bibitem [{\citenamefont {Tse}\ and\ \citenamefont
  {MacDonald}(2010)}]{PhysRevLett.105.057401}%
  \BibitemOpen
  \bibfield  {author} {\bibinfo {author} {\bibfnamefont {Wang-Kong}\
  \bibnamefont {Tse}}\ and\ \bibinfo {author} {\bibfnamefont {A.~H.}\
  \bibnamefont {MacDonald}},\ }\bibfield  {title} {\enquote {\bibinfo {title}
  {Giant magneto-optical kerr effect and universal faraday effect in thin-film
  topological insulators},}\ }\href {\doibase 10.1103/PhysRevLett.105.057401}
  {\bibfield  {journal} {\bibinfo  {journal} {Phys. Rev. Lett.}\ }\textbf
  {\bibinfo {volume} {105}},\ \bibinfo {pages} {057401} (\bibinfo {year}
  {2010})}\BibitemShut {NoStop}%
\bibitem [{\citenamefont {Liang}\ \emph {et~al.}(2021)\citenamefont {Liang},
  \citenamefont {Sukhachov},\ and\ \citenamefont
  {Balatsky}}]{PhysRevLett.126.247202}%
  \BibitemOpen
  \bibfield  {author} {\bibinfo {author} {\bibfnamefont {Long}\ \bibnamefont
  {Liang}}, \bibinfo {author} {\bibfnamefont {P.~O.}\ \bibnamefont
  {Sukhachov}}, \ and\ \bibinfo {author} {\bibfnamefont {A.~V.}\ \bibnamefont
  {Balatsky}},\ }\bibfield  {title} {\enquote {\bibinfo {title} {Axial
  magnetoelectric effect in dirac semimetals},}\ }\href {\doibase
  10.1103/PhysRevLett.126.247202} {\bibfield  {journal} {\bibinfo  {journal}
  {Phys. Rev. Lett.}\ }\textbf {\bibinfo {volume} {126}},\ \bibinfo {pages}
  {247202} (\bibinfo {year} {2021})}\BibitemShut {NoStop}%
\bibitem [{\citenamefont {Kargarian}\ \emph {et~al.}(2015)\citenamefont
  {Kargarian}, \citenamefont {Randeria},\ and\ \citenamefont
  {Trivedi}}]{kargarian2015theory}%
  \BibitemOpen
  \bibfield  {author} {\bibinfo {author} {\bibfnamefont {Mehdi}\ \bibnamefont
  {Kargarian}}, \bibinfo {author} {\bibfnamefont {Mohit}\ \bibnamefont
  {Randeria}}, \ and\ \bibinfo {author} {\bibfnamefont {Nandini}\ \bibnamefont
  {Trivedi}},\ }\bibfield  {title} {\enquote {\bibinfo {title} {Theory of
  {K}err and {F}araday rotations and linear dichroism in topological {W}eyl
  semimetals},}\ }\href@noop {} {\bibfield  {journal} {\bibinfo  {journal}
  {Scientific reports}\ }\textbf {\bibinfo {volume} {5}},\ \bibinfo {pages}
  {1--10} (\bibinfo {year} {2015})}\BibitemShut {NoStop}%
\bibitem [{\citenamefont {Tokman}\ \emph {et~al.}(2020)\citenamefont {Tokman},
  \citenamefont {Chen}, \citenamefont {Shereshevsky}, \citenamefont
  {Pozdnyakova}, \citenamefont {Oladyshkin}, \citenamefont {Tokman},\ and\
  \citenamefont {Belyanin}}]{tokman2020inverse}%
  \BibitemOpen
  \bibfield  {author} {\bibinfo {author} {\bibfnamefont {ID}~\bibnamefont
  {Tokman}}, \bibinfo {author} {\bibfnamefont {Qianfan}\ \bibnamefont {Chen}},
  \bibinfo {author} {\bibfnamefont {IA}~\bibnamefont {Shereshevsky}}, \bibinfo
  {author} {\bibfnamefont {VI}~\bibnamefont {Pozdnyakova}}, \bibinfo {author}
  {\bibfnamefont {Ivan}\ \bibnamefont {Oladyshkin}}, \bibinfo {author}
  {\bibfnamefont {Mikhail}\ \bibnamefont {Tokman}}, \ and\ \bibinfo {author}
  {\bibfnamefont {Alexey}\ \bibnamefont {Belyanin}},\ }\bibfield  {title}
  {\enquote {\bibinfo {title} {Inverse faraday effect in graphene and weyl
  semimetals},}\ }\href@noop {} {\bibfield  {journal} {\bibinfo  {journal}
  {Physical Review B}\ }\textbf {\bibinfo {volume} {101}},\ \bibinfo {pages}
  {174429} (\bibinfo {year} {2020})}\BibitemShut {NoStop}%
\bibitem [{\citenamefont {Sikivie}(1983)}]{Sikivie:1983ip}%
  \BibitemOpen
  \bibfield  {author} {\bibinfo {author} {\bibfnamefont {P.}~\bibnamefont
  {Sikivie}},\ }\bibfield  {title} {\enquote {\bibinfo {title} {{Experimental
  Tests of the Invisible Axion}},}\ }\href {\doibase
  10.1103/PhysRevLett.51.1415} {\bibfield  {journal} {\bibinfo  {journal}
  {Phys. Rev. Lett.}\ }\textbf {\bibinfo {volume} {51}},\ \bibinfo {pages}
  {1415--1417} (\bibinfo {year} {1983})},\ \bibinfo {note} {[Erratum:
  Phys.Rev.Lett. 52, 695 (1984)]}\BibitemShut {NoStop}%
\bibitem [{\citenamefont {Sikivie}(1985)}]{Sikivie:1985yu}%
  \BibitemOpen
  \bibfield  {author} {\bibinfo {author} {\bibfnamefont {Pierre}\ \bibnamefont
  {Sikivie}},\ }\bibfield  {title} {\enquote {\bibinfo {title} {{Detection
  Rates for 'Invisible' Axion Searches}},}\ }\href {\doibase
  10.1103/PhysRevD.36.974} {\bibfield  {journal} {\bibinfo  {journal} {Phys.
  Rev. D}\ }\textbf {\bibinfo {volume} {32}},\ \bibinfo {pages} {2988}
  (\bibinfo {year} {1985})},\ \bibinfo {note} {[Erratum: Phys.Rev.D 36, 974
  (1987)]}\BibitemShut {NoStop}%
\bibitem [{\citenamefont {Raffelt}(1986)}]{Raffelt:1985nk}%
  \BibitemOpen
  \bibfield  {author} {\bibinfo {author} {\bibfnamefont {Georg~G.}\
  \bibnamefont {Raffelt}},\ }\bibfield  {title} {\enquote {\bibinfo {title}
  {{ASTROPHYSICAL AXION BOUNDS DIMINISHED BY SCREENING EFFECTS}},}\ }\href
  {\doibase 10.1103/PhysRevD.33.897} {\bibfield  {journal} {\bibinfo  {journal}
  {Phys. Rev. D}\ }\textbf {\bibinfo {volume} {33}},\ \bibinfo {pages} {897}
  (\bibinfo {year} {1986})}\BibitemShut {NoStop}%
\bibitem [{\citenamefont {Raffelt}\ and\ \citenamefont
  {Stodolsky}(1988)}]{Raffelt:1987im}%
  \BibitemOpen
  \bibfield  {author} {\bibinfo {author} {\bibfnamefont {Georg}\ \bibnamefont
  {Raffelt}}\ and\ \bibinfo {author} {\bibfnamefont {Leo}\ \bibnamefont
  {Stodolsky}},\ }\bibfield  {title} {\enquote {\bibinfo {title} {{Mixing of
  the Photon with Low Mass Particles}},}\ }\href {\doibase
  10.1103/PhysRevD.37.1237} {\bibfield  {journal} {\bibinfo  {journal} {Phys.
  Rev. D}\ }\textbf {\bibinfo {volume} {37}},\ \bibinfo {pages} {1237}
  (\bibinfo {year} {1988})}\BibitemShut {NoStop}%
\bibitem [{\citenamefont {Raffelt}(1990)}]{Raffelt:1990yz}%
  \BibitemOpen
  \bibfield  {author} {\bibinfo {author} {\bibfnamefont {Georg~G.}\
  \bibnamefont {Raffelt}},\ }\bibfield  {title} {\enquote {\bibinfo {title}
  {{Astrophysical methods to constrain axions and other novel particle
  phenomena}},}\ }\href {\doibase 10.1016/0370-1573(90)90054-6} {\bibfield
  {journal} {\bibinfo  {journal} {Phys. Rept.}\ }\textbf {\bibinfo {volume}
  {198}},\ \bibinfo {pages} {1--113} (\bibinfo {year} {1990})}\BibitemShut
  {NoStop}%
\bibitem [{\citenamefont {Gies}(2009)}]{Gies:2008wv}%
  \BibitemOpen
  \bibfield  {author} {\bibinfo {author} {\bibfnamefont {Holger}\ \bibnamefont
  {Gies}},\ }\bibfield  {title} {\enquote {\bibinfo {title} {{Strong laser
  fields as a probe for fundamental physics}},}\ }\href {\doibase
  10.1140/epjd/e2009-00006-0} {\bibfield  {journal} {\bibinfo  {journal} {Eur.
  Phys. J. D}\ }\textbf {\bibinfo {volume} {55}},\ \bibinfo {pages} {311--317}
  (\bibinfo {year} {2009})},\ \Eprint {http://arxiv.org/abs/0812.0668}
  {arXiv:0812.0668 [hep-ph]} \BibitemShut {NoStop}%
\bibitem [{\citenamefont {Redondo}\ and\ \citenamefont
  {Ringwald}(2011)}]{Redondo:2010dp}%
  \BibitemOpen
  \bibfield  {author} {\bibinfo {author} {\bibfnamefont {Javier}\ \bibnamefont
  {Redondo}}\ and\ \bibinfo {author} {\bibfnamefont {Andreas}\ \bibnamefont
  {Ringwald}},\ }\bibfield  {title} {\enquote {\bibinfo {title} {{Light shining
  through walls}},}\ }\href {\doibase 10.1080/00107514.2011.563516} {\bibfield
  {journal} {\bibinfo  {journal} {Contemp. Phys.}\ }\textbf {\bibinfo {volume}
  {52}},\ \bibinfo {pages} {211--236} (\bibinfo {year} {2011})},\ \Eprint
  {http://arxiv.org/abs/1011.3741} {arXiv:1011.3741 [hep-ph]} \BibitemShut
  {NoStop}%
\end{thebibliography}%

\end{document}